\documentclass[screen,nonacm,sigplan,10pt]{acmart}
\renewcommand\footnotetextcopyrightpermission[1]{}
\usepackage{amsmath,amsfonts}
\usepackage{algorithm}
\usepackage{algorithmic}
\usepackage{graphicx}
\usepackage{textcomp}
\usepackage{xcolor}
\usepackage[]{hyperref}
\usepackage{cleveref}
\usepackage{mathtools}
\usepackage{xspace}
\usepackage{graphicx}
\usepackage{caption}
\usepackage{subcaption} 
\usepackage{enumitem}
\usepackage{multirow}
\usepackage{makecell}
\usepackage{kotex}
\usepackage{comment}
\usepackage{subcaption}
\usepackage{subfiles}
\usepackage{listings}
\usepackage[frozencache,cachedir=.]{minted}
\usepackage{mdframed}
\usepackage[most]{tcolorbox}
\usepackage{enumitem}

\lstdefinestyle{codestyle}{
    basicstyle=\ttfamily\scriptsize,
    breaklines=true,
    numbers=left,
    numberstyle=\tiny,
    frame=single,
    columns=fullflexible,
    keepspaces=true,
    showstringspaces=false,
    tabsize=2
}

\newcommand{\C}[1]{{\color{blue}#1}}

\newcommand{\name}{FS-SSD\xspace}
\newcommand{\proto}{NFS-over-NVMe\xspace}

\newcommand{\dw}{\texttt{dword}\xspace}
\newcommand{\dws}{\texttt{dwords}\xspace}
\newcommand{\dwn}[1]{\texttt{dword{#1}}\xspace}

\newlist{requirements}{description}{1}
\setlist[requirements]{
leftmargin=3.2em,
labelwidth=2.6em,
labelsep=0.6em,
itemsep=0.35em,
topsep=0.25em,
font=\normalfont\bfseries
}

\AtBeginDocument{%
  }

\setcopyright{acmlicensed}
\copyrightyear{2026}
\acmYear{2026}
\acmDOI{XXXXXXX.XXXXXXX}
\acmConference[Conference acronym 'XX]{Make sure to enter the correct
  conference title from your rights confirmation email}{June 03--05,
  2018}{Woodstock, NY}
\acmISBN{978-1-4503-XXXX-X/2018/06}

\begin{document}
\pagestyle{plain}
\title{File-Semantic SSD for I/O-centric Architectures with \textit{NFS-over-NVMe} Protocol}


\author{Kyu-Jin Cho}
\affiliation{%
  \institution{Seoul National University}
  \city{Seoul}
  \country{Republic of Korea}}
\email{bori19960@snu.ac.kr}

\author{Injae Kang}
\affiliation{%
  \institution{Seoul National University}
  \city{Seoul}
  \country{Republic of Korea}}
\email{abcinje@snu.ac.kr}

\author{Keunsan Park}
\affiliation{%
  \institution{Seoul National University}
  \city{Seoul}
  \country{Republic of Korea}}
\email{keunsan.park@snu.ac.kr}

\author{Jaehoon Shim}
\affiliation{%
  \institution{Seoul National University}
  \city{Seoul}
  \country{Republic of Korea}}
\email{mattjs@snu.ac.kr}

\author{Munsung Bae}
\affiliation{%
  \institution{Seoul National University}
  \city{Seoul}
  \country{Republic of Korea}}
\email{hamasoon@snu.ac.kr}

\author{Jiho Jeon}
\affiliation{%
  \institution{Seoul National University}
  \city{Seoul}
  \country{Republic of Korea}}
\email{jeonjiho9030@snu.ac.kr}

\author{Eyee Hyun Nam}
\affiliation{%
  \institution{FADU Inc.}
  \city{Seoul}
  \country{Republic of Korea}}
\email{ehnam@fadu.io}

\author{Jin-Soo Kim}
\affiliation{%
  \institution{Seoul National University}
  \city{Seoul}
  \country{Republic of Korea\\}
  \institution{FADU Inc.}
  \city{Seoul}
  \country{Republic of Korea}}
\email{jinsoo.kim@snu.ac.kr}


\begin{abstract}
Emerging I/O-centric architectures empower peripheral devices to access storage directly while bypassing the host CPU and kernel. 
However, SSDs expose only a block interface and rely on the host filesystem for file metadata and access control, making file-level access for applications and peripheral devices inherently difficult in I/O-centric architectures.
This tight coupling between the host filesystem and storage device introduces unavoidable software overhead and limits the potential of fully hostless I/O paths.

In this paper, we present File-Semantic SSD (\name), a new SSD architecture that provides a native file abstraction within SSDs, enabling direct file access by peripheral and remote devices without host intervention. \name adopts the lightweight, stateless NFS interface and introduces \proto, a protocol extension that embeds NFS operations into the standard NVMe command structure. This allows \name to support local and remote file I/O through existing PCIe and NVMe-oF transports.
Our evaluation shows that \name{} provides on-device file abstraction while fully utilizing raw-device bandwidth in I/O-centric architectures.
\end{abstract}

\maketitle

\section{Introduction}

With the exponential growth of data volumes and workload demands, modern servers and storage appliances are increasingly equipped with high-performance peripheral devices, including high-bandwidth I/O components~\cite{ssd,nic,smartnic}, non-volatile memory (NVM)~\cite{nvm,optane}, and hardware accelerators~\cite{a100,h100,fpga}.
The performance of these devices has continued to advance, largely driven by innovations in interconnect technologies such as PCIe~\cite{pcie}, InfiniBand~\cite{infiniband}, and NVLINK~\cite{nvlink}, which ensure rapid data transfer between components.
For example, today's storage devices can deliver bandwidth of up to tens of GB/s~\cite{ssd,pm1763}, while offering access latencies as low as a few microseconds~\cite{ull-ssd}.

However, as device performance and interconnect speeds have improved, the overhead introduced by the OS kernel has become increasingly pronounced. 
Numerous studies have shown that the Linux I/O stack can significantly degrade performance due to factors such as context switches~\cite{asio, xrp}, interrupt handling~\cite{nvmedirect, spdk, pollinterrupt, osio}, file-to-block translation~\cite{strata, splitfs, zofs}, and network stack~\cite{io-tcp, understanding}.
In some cases, this overhead can account for more than 50\% of total execution time~\cite{bpfforstorage, dcs-ctrl,bypassd}. 
As a result, the kernel software stack has emerged as a major performance bottleneck.

To mitigate such software overheads, \textit{I/O-centric architectures}~\cite{spdk, bam, dcs-ctrl} have been introduced to treat I/O devices as first-class citizens.
These architectures minimize CPU involvement in the I/O path by enabling both CPU applications and peripheral devices, such as GPUs, SmartNICs, and DPUs, to directly initiate and handle I/O operations without OS kernel intervention.
System designs adopting I/O-centric architectures have been proposed across a wide range of domains, including machine learning~\cite{gds, bam, gids, hyperion, gmt, neos}, networking~\cite{io-tcp, dcs-ctrl}, and in-storage processing~\cite{presto, futurecloud, bluedbm, summarizer, insider}.


While I/O-centric architectures are gaining traction, the lack of native file abstraction within storage devices makes it difficult to build and integrate applications on top of these architectures.
Traditionally, applications access storage devices through the host filesystem, which resolves pathnames, enforces access control, locates data blocks, and manages metadata.
File semantics therefore not only provide a convenient means of naming data, but also define the abstraction through which applications locate, protect, and share persistent data.
As long as these semantics remain confined to the host, this tight coupling becomes a critical bottleneck when applications and peripheral devices attempt to access file data directly or offload computation to the storage. 
In other words, I/O-centric architectures must either give up file semantics to avoid CPU involvement or repeatedly call back into the host filesystem to preserve them.

In this paper, we present File-Semantic SSD (\name), a novel SSD architecture that provides file abstraction directly within the storage device for I/O-centric architectures.
With \name, applications and peripheral devices can access files directly over high-performance I/O channels without relying on the host-side filesystem. 
Specifically, \name adopts the lightweight, industry-proven Network File System (NFS) interface, which is well-suited for resource-constrained SSDs due to its simplicity and minimal runtime state.
To support this model, we introduce a new protocol called \textit{\proto}, which defines a custom command set that integrates NFS operations into the standard NVMe command layout. 
This design enables other devices to access \name using a familiar file interface built directly on the NVMe protocol. 
Because \proto adheres to the NVMe specification~\cite{nvme}, \name can natively support advanced features such as NVMe-over-Fabric Target Offload~\cite{targetoffload}, enabling efficient remote file access without involving the host CPU.

To efficiently support \proto, \name integrates a co-designed file layer and Flash Translation Layer (FTL) based on the flash-friendly F2FS filesystem~\cite{f2fs}.
In particular, \name is carefully engineered to operate within the same hardware constraints as conventional SSDs, including limited DRAM and compute resources, making it practical for real-world deployment.
Furthermore, it is designed to provide the file abstraction within the storage device without compromising the available bandwidth of the SSD.

We conducted a series of experiments in both local and remote environments to evaluate the effectiveness of \name in I/O-centric architectures.
Our evaluation in local environments shows that \name provides a native on-device file abstraction while fully utilizing raw-device bandwidth, and improves file I/O throughput for I/O-centric architectures.
In remote environments, we demonstrate that \name outperforms NFS implementations while eliminating target-side CPU involvement. 
Specifically, under CPU-constrained conditions on the storage server, \name delivers up to 4.2x higher throughput compared to NFS-over-RDMA.
Experiments on real-world GPU workloads including graph analytics and LLM training further demonstrate that \name can provide a robust file abstraction for I/O-centric architectures. 
The source code of \name will be made available at GitHub.

\section{Background and Motivation}\label{background}

\subsection{NVMe SSD} 
Modern Solid State Drives (SSDs) are built on NAND flash memory and internally manage data using a hierarchy of physical units. 
The smallest unit of read and write is a \textit{flash page}, which typically includes both a data region and a small out-of-band (OOB) area for metadata and error correction. 
Pages are grouped into \textit{flash blocks}, the smallest erasable unit, and further aggregated into \textit{superblocks} to exploit internal parallelism across multiple flash channels and dies.

To present a conventional block device abstraction to the host, SSDs rely on a firmware component known as the \textit{Flash Translation Layer (FTL)}. 
The FTL maintains a logical-to-physical (L2P) mapping table that translates logical block addresses (LBAs) into physical page addresses (PPAs) in flash memory.
Because flash pages cannot be updated in place, the FTL also performs \textit{FTL-level garbage collection (GC)}, which reclaims space by erasing blocks containing obsolete data after relocating any valid pages.
These functions are supported by multiple embedded CPU cores within the SSD controller and several gigabytes of onboard DRAM which typically accounts for about 0.1\% of the total storage capacity~\cite{970pro,990pro}.

Communication between the host and SSD is governed by the NVMe interface~\cite{nvme}, which is designed to exploit PCIe's high-bandwidth, low-latency characteristics.
NVMe employs a set of paired submission and completion queues (SQs and CQs) that support parallel I/O across CPU cores.
The host issues I/O commands into SQs and signals the device by writing to a doorbell register.
After executing the commands, the SSD posts completion entries to the corresponding CQs and the host is notified via interrupts or may poll the CQs to retrieve the completion entries.

Beyond local storage, NVMe has been extended through NVMe-over-Fabrics (NVMe-oF)~\cite{nvme-of}, a protocol that enables NVMe commands to be transmitted over network transports such as RDMA or TCP.
NVMe-oF allows remote hosts to access SSDs as if they were locally attached, facilitating the deployment of high-performance disaggregated storage systems in modern data centers.
With \textit{Target Offload} capabilities~\cite{targetoffload}, advanced NICs can process NVMe-oF requests entirely in hardware, bypassing the remote CPU and further reducing both I/O latency and host overhead. 

\subsection{Network File System}
Network File System (NFS)~\cite{nfs} is a widely adopted remote filesystem that allows clients to access files over the network.
It follows a client-server architecture, where the server exports its files and directories and clients mount those directories into their local filesystem namespace.
Through the NFS protocol which is based on Remote Procedure Calls (RPCs), clients can read, write, and modify remote files transparently, as if they were stored locally.
To represent files and directories, NFS uses a compact and persistent identifier known as a \textit{file handle}.
When a client first accesses a file via its pathname, the server resolves it and returns a corresponding file handle. 
Subsequent operations use this handle directly, allowing the server to avoid repeated path resolution.

To reduce software overhead, NFS-over-RDMA (NFS-oR) \cite{nfs-or} extends the traditional NFS protocol by leveraging RDMA transports instead of TCP/IP.
This enables direct memory-to-memory transfers between client and server, eliminating intermediate data copies and significantly lowering CPU utilization.
However, despite reducing networking overhead, NFS-oR still relies on the host kernel and CPU to process RPCs and manage filesystem semantics, which can remain a performance bottleneck in I/O-intensive workloads.

\begin{figure}[t]
    \centering
        \includegraphics[width=\columnwidth]{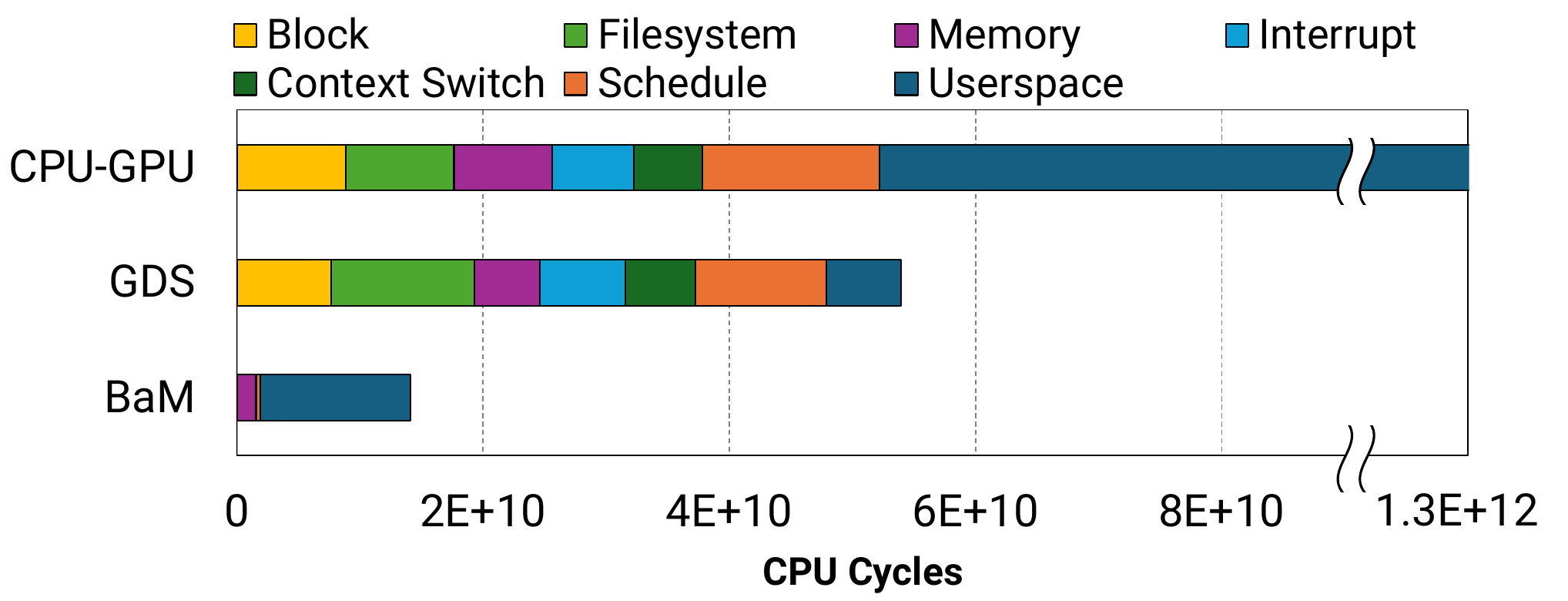} 
    \vspace{-0.2cm}
    \caption{\textbf{CPU cycle breakdown for three I/O scenarios between a GPU and an SSD.}}
    \label{fig:M2} 
    \vspace{-0.5cm}
\end{figure}

\subsection{I/O-centric Architectures}\label{io-centric}
I/O-centric architecture refers to a system design approach that aims to minimize or eliminate CPU involvement in the I/O path, particularly for I/O-intensive workloads.
Instead of routing all I/O operations through the kernel on the host CPU, it enables peripheral devices to initiate, process, and complete data transfers independently.
By offloading both control- and data-plane responsibilities to the devices themselves, I/O-centric architectures reduce software overhead, enhance I/O efficiency, and unlock the full potential of modern high-performance hardware.

This architectural model is becoming increasingly prevalent with the emergence of peripheral devices that offer significant computational capabilities.
IO-TCP~\cite{io-tcp}, for instance, enables high-performance data delivery for streaming services by offloading both TCP/IP processing and storage I/O to a DPU~\cite{smartnic}.
GIDS~\cite{gids} and Hyperion~\cite{hyperion} leverage GPUDirect Storage (GDS)~\cite{gds} and GPU-initiated I/O~\cite{bam, scada}, respectively, to accelerate Graph Neural Network (GNN) training by transferring data directly to GPU memory.
PreSto~\cite{presto} introduces in-storage data preprocessing for recommendation model training, while Summarizer~\cite{summarizer} offloads portions of application logic to computational storage devices (CSDs), enabling data filtering, transformation, and analysis to occur near the storage media.
These systems highlight the growing shift toward I/O-centric architectures, where I/O devices take over responsibilities traditionally handled by the host.

While such structural optimizations toward I/O-centric architectures are highly effective at reducing CPU involvement, they introduce a tradeoff between performance and usability: users must either operate on raw data locations or fall back to the host filesystem to locate data.
To quantitatively assess this tradeoff, we measure the CPU cycle usage across three I/O scenarios involving 4\,KB data transfers between a GPU and an SSD. 
Specifically, we compare three approaches: (1) CPU-GPU that transfers data from storage through the host CPU using \texttt{cudaMemcpy()}, (2) GDS~\cite{gds} and (3) BaM~\cite{bam}.
GDS establishes a direct data path for DMA transfers between GPU and storage, eliminating the need for a CPU-side bounce buffer.
In contrast, BaM allows a GPU to directly initiate I/O requests without host CPU intervention to access data stored in storage devices.

\Cref{fig:M2} presents the CPU cycle breakdown for each scenario, highlighting the layers responsible for CPU consumption.
We observe that the CPU-GPU approach incurs the highest CPU overhead, with approximately 95\% of CPU cycles spent by \texttt{cudaMemcpy()} in userspace.
In contrast, GDS significantly reduces userspace CPU usage by enabling the CPU to directly transfer data from storage to GPU memory.
However, the filesystem and block layers still account for around 30\% of total CPU cycles. 
This is because the GPU lacks awareness of the physical block locations on the SSD, and thus file-to-block translation must still be performed by the host CPU before I/O requests can be issued.
Moreover, context switches and interrupt handling also consume a non-negligible share of CPU cycles, as the GDS control plane still depends on the host CPU to initiate I/O operations to the GPU and handle their completions. 

BaM demonstrates the lowest CPU usage among all evaluated approaches.
This is attributed to BaM's design, which allows the GPU to manage both the submission and completion of I/O requests, fully bypassing the kernel.
However, BaM lacks support for file abstraction, requiring users to manually determine data block locations in advance, posing a usability challenge for general-purpose applications.
Users must preload all datasets at predefined storage offsets before launching applications on BaM, and maintain corresponding location metadata in GPU memory during execution.
This not only degrades the usability of BaM, but also limits the portability of applications and file sharing.
These limitations are fundamental to all I/O-centric architectures that bypass the host kernel I/O stack, as they cannot independently resolve the exact data locations within the storage device.

\begin{figure*}[ht]
    \centering
    \begin{subfigure}[ht]{0.17\textwidth} 
        \centering
        \includegraphics[width=\textwidth]{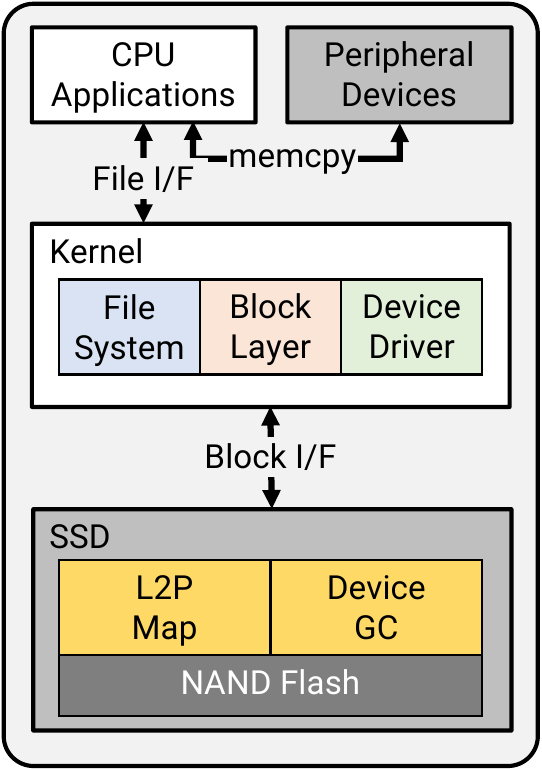} 
        \caption{Traditional FS} 
        \label{fig:M1-1}
    \end{subfigure}
    \hspace{0.2cm}
    \begin{subfigure}[ht]{0.17\textwidth}
        \centering
        \includegraphics[width=\textwidth]{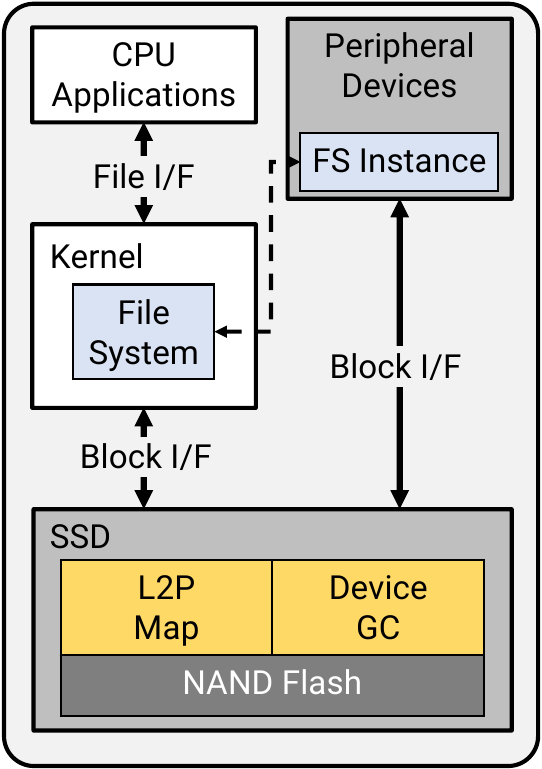} 
        \caption{Coordinated FS} 
        \label{fig:M1-2}
    \end{subfigure}
    \hspace{0.2cm}
    \begin{subfigure}[ht]{0.17\textwidth}
        \centering
        \includegraphics[width=\textwidth]{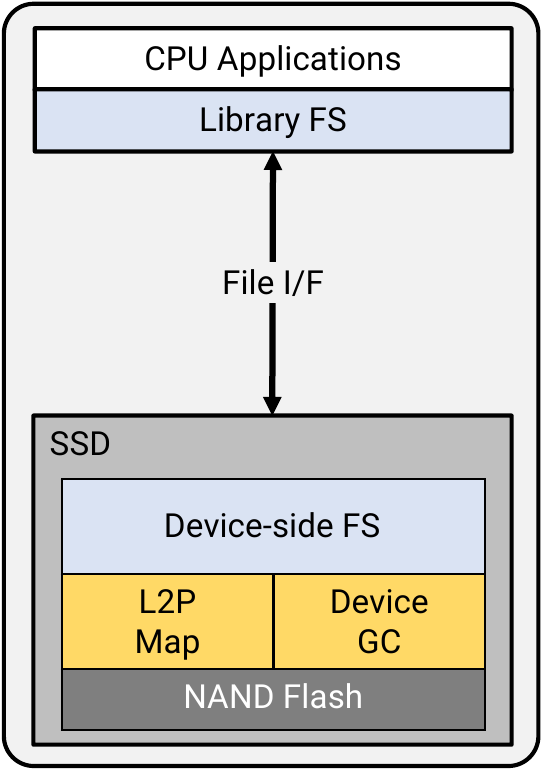} 
        \caption{Direct-Access FS} 
        \label{fig:M1-3}
    \end{subfigure}
    \hspace{0.2cm}
    \begin{subfigure}[ht]{0.17\textwidth}
        \centering
        \includegraphics[width=\textwidth]{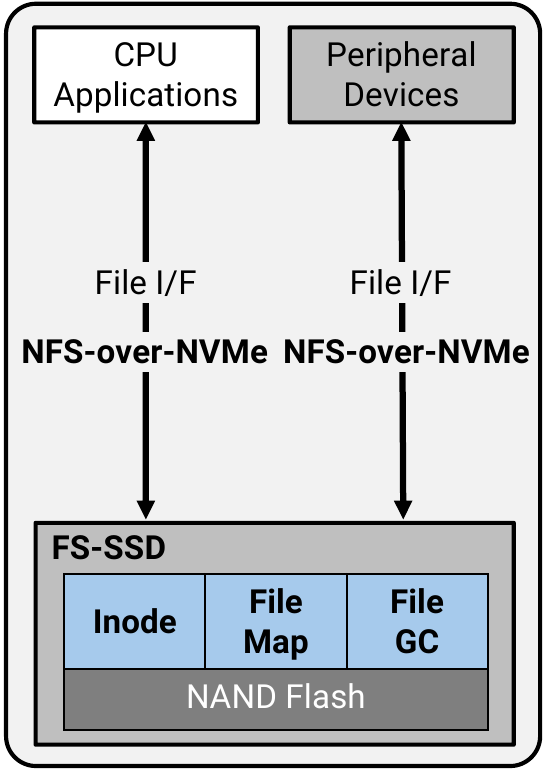} 
        \caption{\name (Local)} 
        \label{fig:M1-4}
    \end{subfigure}
    \hspace{0.2cm}
    \begin{subfigure}[ht]{0.17\textwidth}
        \centering
        \includegraphics[width=\textwidth]{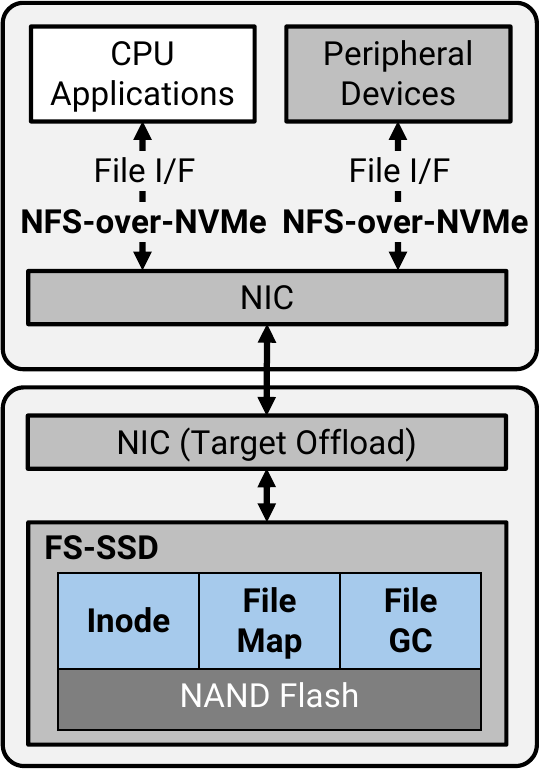} 
        \caption{\name (Remote)} 
        \label{fig:M1-5}
    \end{subfigure}
    \caption{\textbf{Providing file abstraction for SSDs.}}
    \label{fig:M1}
\end{figure*}

\subsection{File Abstraction for Storage Devices}
In the era of I/O-centric architectures, applications built on these architectures face a unique challenge: file semantics reside in the host kernel, not on the storage device itself.
SSDs expose only a block-level interface, relying on the host filesystem to interpret raw blocks as files and locate the underlying data, as shown in~\Cref{fig:M1-1}. 
This tight coupling between the host filesystem and the storage device becomes a major obstacle when applications and other peripheral devices attempt to directly access data stored on SSDs or coordinate in-storage computations.

Several approaches have been proposed to address the absence of native file abstraction in storage devices.
Rather than issuing filesystem requests to the host on every access, one straightforward approach is to extract file extent metadata from the host filesystem via the \texttt{ioctl(FIEMAP)} system call and deliver this information to the device in advance~\cite{insider, lambda-io, aisio}. 
However, approaches that rely on fixed, pre-extracted mappings require file space to be preallocated, restricting dynamic file growth and flexibility.
Moreover, mappings extracted before application execution may become stale as the filesystem state changes at runtime, potentially compromising consistency and allowing unintended access to reallocated blocks.

A more structured solution is the \textit{Coordinate Filesystem} approach~\cite{gpufs,lightfs,geminifs,dockerssd,gofs}, in which a simplified filesystem instance runs directly on the I/O device.
As illustrated in~\Cref{fig:M1-2}, the device operates on a shadow copy of the host filesystem's metadata, allowing it to perform file-level operations without involving the kernel on every access.
This decoupling reduces kernel interactions during data path operations, thereby improving I/O performance.
However, ensuring consistent views between the host and device-side instances is non-trivial. 
Because updates must be synchronized across both views, the system incurs coordination overhead and frequent round-trips to the host, ultimately limiting the potential for full kernel bypass. 
As a variant, Newport CSD~\cite{newportcsd} proposes synchronizing host and device filesystems using a cluster filesystem over TCP/IP tunneling through the PCIe bus, which introduces additional overhead. 


Finally, \textit{In-storage Filesystems} such as DevFS~\cite{devfs} and CrossFS~\cite{crossfs} implement the filesystem logic inside the storage device firmware and expose traditional file interfaces as depicted in~\Cref{fig:M1-3}.
While these systems can offer consistent namespace and low-latency, direct access to storage, their architectures are difficult to realize within resource budgets of commercial SSDs since they independently stack a device-side filesystem layer on top of SSD's FTL.
Moreover, they are originally designed as local filesystems for CPU-based user applications, and thus do not natively support access from other I/O devices or remote access scenarios. 
Another approach with a motivation similar to ours is NFS-eSSD~\cite{nfs-essd}, which integrates NFS server functionality directly within an SSD. 
However, NFS-eSSD requires both filesystem logic and an on-device network stack to process NFS requests.
Implementing these components within an SSD's resource constraints requires careful vertical co-design of the network stack, filesystem, and FTL.
In summary, these systems leave room for further optimization through comprehensive co-design with the underlying FTL, accounting for device-level resource constraints to make more efficient use of limited on-device DRAM and processing resources.

\subsection{Motivation}

%


The lack of native file abstraction in storage devices remains a fundamental barrier to realizing fully hostless storage access paths. 
This work aims to address challenges by developing a resource-efficient in-storage file layer that enables standalone file-level access entirely within the storage device. 
The proposed approach must satisfy the following key requirements tailored for I/O-centric architectures: 

\noindent \textbf{R1}: The file interface should be simple and lightweight, minimizing overhead for both developers and hardware.

\noindent \textbf{R2}: File data must be directly accessible by other I/O devices, including remote devices, without host-CPU involvement.

\noindent \textbf{R3}: Files must be sharable between the host CPU and I/O devices with well-defined access control and consistency semantics to ensure correctness.

\noindent \textbf{R4}: The system should support fine-grained concurrency and crash consistency, enabling robust operation in multi-threaded and failure-prone environments.

\noindent \textbf{R5}: The system should be integrated with the host VFS layer seamlessly, preserving compatibility while delivering performance comparable to conventional host-based filesystems.

\noindent \textbf{R6}: The system must be resource-efficient, capable of operating within the limited DRAM and compute resources typically available on SSDs.

In this paper, we propose File-Semantic SSD (\name) designed to meet the aforementioned requirements.
To satisfy \textbf{R1}, \textbf{R3}, and \textbf{R5}, \name adopts the NFS version 3 (NFSv3) interface, which is a widely deployed, industry-proven technology. 
NFSv3 is designed for simplicity and speed, requiring only 21 operations. 
It naturally supports multi-client file sharing semantics and is well-integrated with the Linux software stack, allowing us to leverage existing NFS client and mount utilities. 
To meet \textbf{R2}, we introduce \textit{\proto}, a new protocol that carries NFS commands over the standard NVMe interface, enabling other PCIe-based I/O devices to access files on \name directly without CPU involvement, as shown in~\Cref{fig:M1-4}.
Furthermore, since NVMe-oF is widely supported by modern NICs and DPUs, \proto can be extended to remote environments without introducing additional protocol layers, as shown in \Cref{fig:M1-5}.
Lastly, to satisfy \textbf{R4} and \textbf{R6}, \name leverages design principles of F2FS~\cite{f2fs}, a flash-friendly, log-structured filesystem. 
Since F2FS's metadata layout and garbage collection mechanisms are well aligned with the behavior of the FTL, we co-design the filesystem and FTL layers to unify their responsibilities, thereby enabling resource-efficient operation within the constrained DRAM and compute environment of SSDs.

\section{\proto Protocol}\label{nfsonvme}

%

\subsection{Why NFSv3?}

Our design adopts NFS version 3 (NFSv3) as the foundation of the \proto protocol.
Several factors make NFSv3 particularly suitable for our design:

\noindent\textbf{$\bullet$ Lightweight Interface:} NFSv3 defines a minimal set of RPCs that cover essential file operations. This streamlined operation set reduces both implementation complexity and overhead in resource-constrained environments like SSDs.

\noindent\textbf{$\bullet$ OS- and Filesystem-Agnostic:} NFSv3 is independent of the underlying OS and filesystem, making it well-suited for heterogeneous systems and file sharing. 

\noindent\textbf{$\bullet$ Stateless Semantics:} NFSv3 is designed to be stateless, relying on idempotent operations that minimize the amount of state that must be tracked by the server. This property is especially advantageous for firmware-level implementations, where persistent state management is costly and error-prone.

\subsection{Command Layout}
\begin{figure}[t]
    \centering
    \begin{subfigure}[ht]{0.45\columnwidth} 
        \centering
        \includegraphics[width=\textwidth]{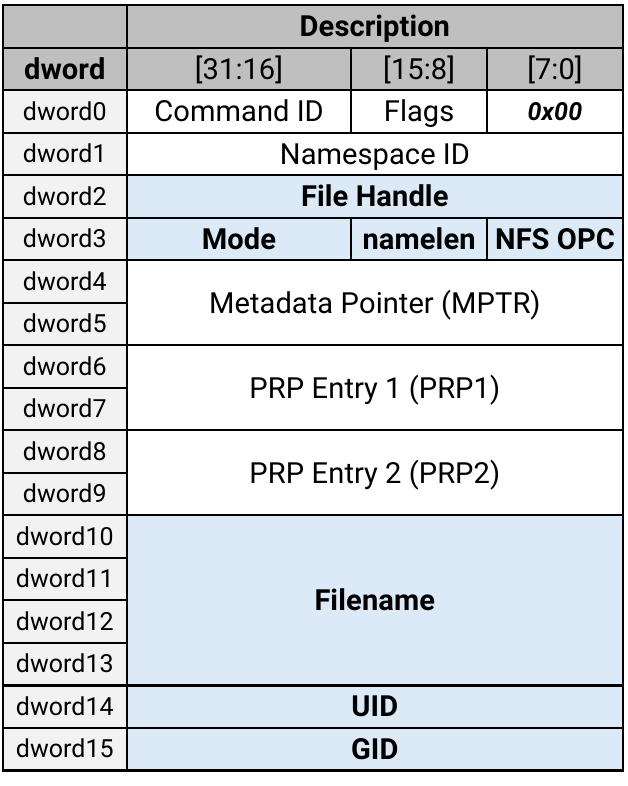} 
        \caption{Category: Normal} 
        \label{fig:D1-1}
    \end{subfigure}
    \hfill
    \begin{subfigure}[ht]{0.45\columnwidth}
        \centering
        \includegraphics[width=\textwidth]{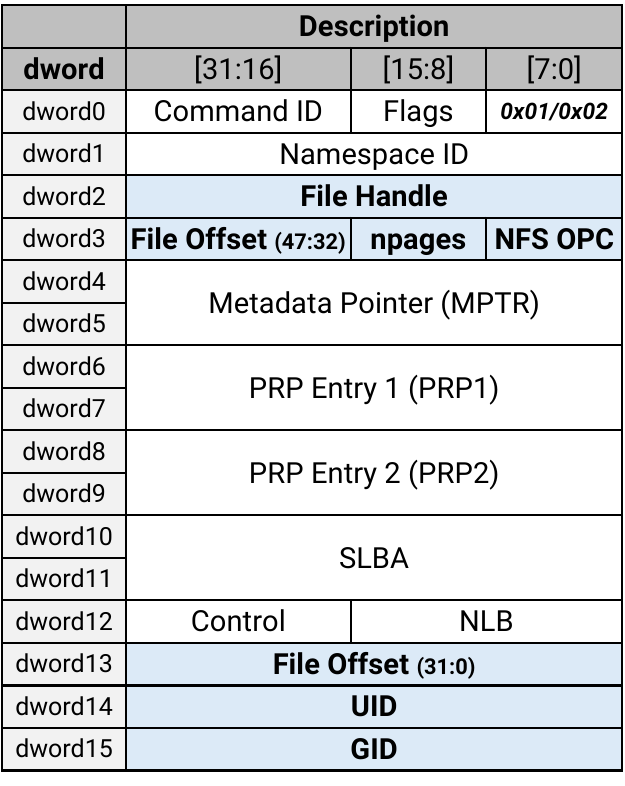} 
        \caption{Category: A2F/F2A} 
        \label{fig:D1-2}
    \end{subfigure}
    \vspace{-0.2cm}
    \caption{\textbf{\proto command layout of each category.} The opcodes for NVMe \textit{flush}, \textit{write} and \textit{read} commands are \texttt{0x00}, \texttt{0x01}, and \texttt{0x02} in \dwn{0}, respectively.} 
    \vspace{-0.4cm}
    \label{fig:D1}
\end{figure}

A standard NVMe command consists of 16 double words (\dws), with each \dw being 4 bytes in size.
The layout and usage of these \dws are defined by the NVMe specification~\cite{nvme}.
For example, the first byte of \dwn{0} contains an NVMe opcode, while \dwn{6-9} designate Physical Region Page (PRP) entries, which provide the physical memory addresses for data buffers (see~\Cref{fig:D1-1}).
We define \proto commands by extending the standard NVMe command set to support NFS operations and repurposing specific \dws to encode file-related parameters. 
Each NFS operation has its own NFS opcode (NFS OPC) to uniquely identify it as part of the \proto protocol and it is assigned to a specific NVMe opcode. 
Additionally, we categorize \proto commands based on whether they require additional source or destination data buffers between the client and \name.
\Cref{Table:D1} summarizes this categorization and \Cref{fig:D1} illustrates the corresponding command layout for each operation category.
Note that these commands are distinguished from standard NVMe commands by setting a reserved \textit{flags} bit in \dwn{0}, allowing \name to coexist seamlessly with conventional SSDs.

\begin{table}[t]
\caption{\textbf{Categories of core \proto commands.}}
\vspace{-0.2cm}
\resizebox{\columnwidth}{!}{%
\begin{tabular}{c|c|c}
\Xhline{3\arrayrulewidth}
\textbf{Category }      & \textbf{NVMe Command}      & \textbf{NFS Operations}                                                                \\ \Xhline{3\arrayrulewidth}
\multirow{2}{*}{Normal} & \multirow{2}{*}{Flush}     & \textsc{create}, \textsc{mkdir}, \textsc{mknod}, \textsc{lookup},     \\ 
                        &                            & \textsc{access}, \textsc{remove}, \textsc{rmdir}, \textsc{commit}                      \\ \hline
\multirow{1}{*}{A2F}    & \multirow{1}{*}{Write}     & \textsc{write}, \textsc{setattr}, \textsc{symlink}, \textsc{link}, \textsc{rename}     \\ \hline
\multirow{2}{*}{F2A}    & \multirow{2}{*}{Read}      & \textsc{read}, \textsc{readdir}, \textsc{readdirplus}, \textsc{readlink},              \\
                        &                            & \textsc{getattr}, \textsc{fsstat}, \textsc{fsinfo}, \textsc{pathconf}                  \\ \Xhline{3\arrayrulewidth}
\end{tabular}%
}
\vspace{-0.3cm}
\label{Table:D1}
\end{table}

\noindent\ \ (1) \textbf{\textit{Normal}}: 
This category includes NFS operations that can encapsulate all required parameters directly within a single NVMe command, and whose results fit within a Completion Queue Entry (CQE). 
As shown in \Cref{fig:D1-1}, these operations extend the NVMe \textit{flush} command by embedding the parameters specific to each NFS operation into the command structure.
For example, a \textsc{create} operation places parameters such as the \textit{parent directory handle} and \textit{mode} into \dwn{2} and \dwn{3}, and the filename in \dwn{10-13}. 
%

%
\noindent\ \ (2) \textbf{\textit{A2F}} (Application to Filesystem): 
This category includes operations that require transferring data from the application to \name. 
It utilizes the NVMe \textit{write} command to transmit the address of the data buffer.
Unlike traditional NVMe \textit{write} commands that specify block-level information, A2F operations embed file-level properties, such as the file handle and offset, directly into the command, as illustrated in~\Cref{fig:D1-2}. 
This mechanism is also used for operations whose parameters exceed the space available in a single NVMe command. 
For example, the \textsc{setattr} operation places the file attributes in a buffer and specifies its location using PRP entries.
Additionally, if the filename in a Normal operation exceeds 16 bytes (i.e., the capacity of 4 \dws), an extra buffer is allocated to hold the filename, and the request is sent using the same mechanism as an A2F operation.

\noindent\ \ (3) \textbf{\textit{F2A}} (Filesystem to Application): 
In contrast to A2F, this category includes operations that require a buffer to receive results from \name. 
F2A operations use the same command layout as
A2F, but are issued using the NVMe \textit{read} command along with PRP entries that point to the destination buffer.
This layout is also used for operations whose results exceed the size of a CQE, such as \textsc{readdir} and \textsc{getattr}.

\subsection{Remote Storage Support}
%
The \proto protocol is carefully designed to remain compliant with the NVMe specification, enabling \name to operate as a remote storage target over the NVMe-oF protocol.
Using this protocol, \proto commands can be transmitted from the client's NIC to the target server via RDMA.
The target server then forwards these commands to the \name device. 
Furthermore, our design supports the Target Offload~\cite{targetoffload} feature, which allows the target server's NIC to bypass the target-side host OS and deliver commands directly to \name.
Unlike traditional NFS, which routes RPC requests to an NFS server running on the CPU, this path eliminates the server CPU from the remote file-access path.
As a result, \name achieves remote file-access performance comparable to traditional block access over NVMe-oF.

\begin{figure}[t]
    \centering
        \includegraphics[width=\columnwidth]{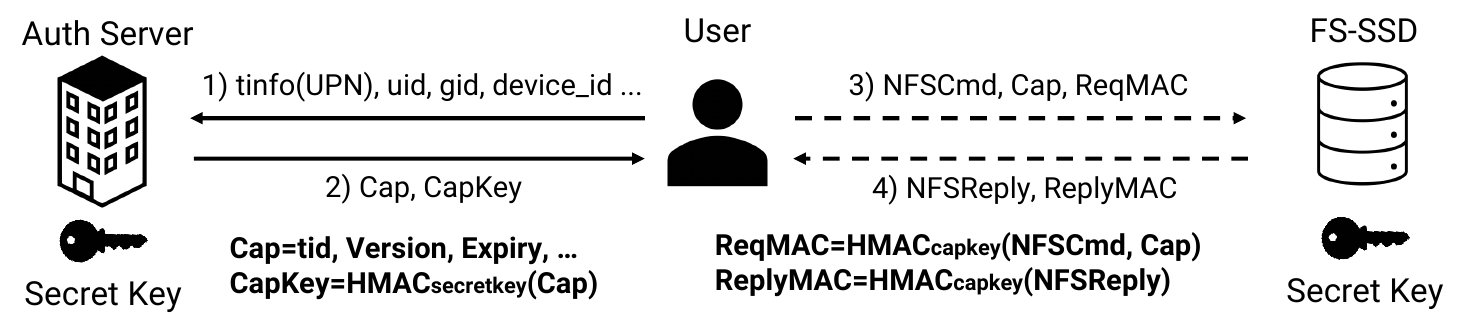} 
    \vspace{-0.6cm}
    \caption{\textbf{Capability-based secure communication.}}
    \vspace{-0.4cm}
    \label{fig:D4} 
\end{figure}

\subsection{Capability-based Security}
%
In local deployments, \name trusts on-node components such as the host kernel and attached peripheral devices.
For remote deployments, it authenticates clients through the unique \textit{hostnqn} identifier~\cite{nvme} defined by the NVMe protocol.
Thus, only clients pre‑authenticated by administrators can connect to a remote \name.
However, this is not sufficient in multi-tenant environments, where a single host serves multiple mutually untrusted users.
Thus, \name provides capability-based security, which is prevalent in networked storage systems~\cite{nasd1, cephx, seaweedfs, s3presignedurl}.

\Cref{fig:D4} illustrates how \name establishes secure communication with a user.
First, the user requests an access capability from the authentication server (Auth Server) over a secure channel external to \name by submitting tenant information including the User Principal Name (UPN).
The server then verifies user's identity and returns a capability (Cap) and a CapKey if the request is granted. 
The capability encodes the tenant identifier (tid) and the access rights in \name, while CapKey is a cryptographic token derived from a secret key shared only with \name.
When the user subsequently issues \proto commands, it computes a ReqMAC over the command and Cap using CapKey, and sends them along with each command.
Since \name shares the secret key with the Auth Server, it can re-derive CapKey from the received Cap and verify the attached ReqMAC. 
If the recomputed ReqMAC matches the received one, \name then checks the tenant’s identity and node's permission fields to determine whether the tenant is authorized to access the target files.
Any modification to these fields causes ReqMAC verification to fail, and \name rejects the request with an access-denied error.
The same mechanism is applied to ReplyMAC to ensure that the reply is not modified in transit.
It guarantees that every request is properly authorized and that both requests and replies cannot be forged.

\section{\name Design}\label{design}
\subsection{Overall Architecture}

\begin{figure}[t]
    \centering
        \includegraphics[width=0.7\columnwidth]{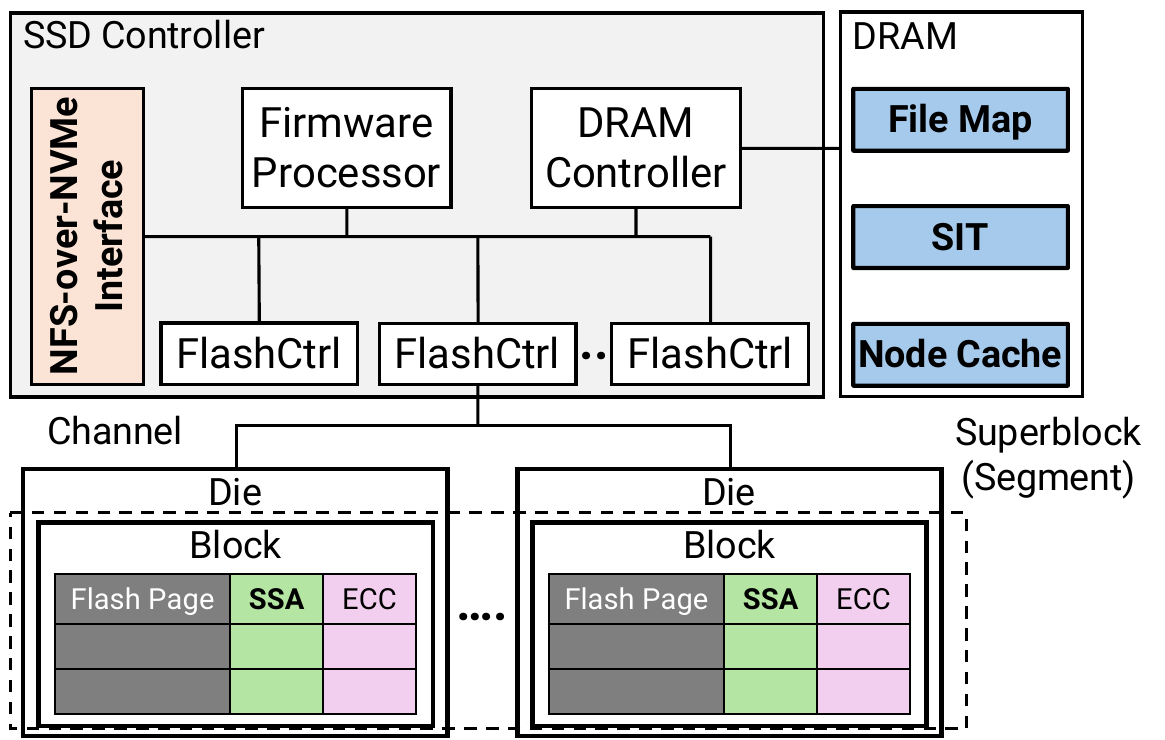} 
    \caption{\textbf{Overall architecture of \name.}} 
    \vspace{-0.4cm}
    \label{fig:D2} 
\end{figure}

As \name operates directly on raw storage hardware, it is designed to leverage low-level device internals to optimize both data and metadata management.
Specifically, \name is built around the core filesystem structures of F2FS~\cite{f2fs}, a flash-optimized, log-structured file system. 
F2FS organizes all writes in an append-only manner, which is particularly well-suited to the characteristics of NAND flash.

F2FS manages file data and metadata using two types of 4~KB blocks: data blocks and node blocks.
Each node block is identified by a unique \textit{node ID} and can represent one of three types: \textit{inode} nodes, which store file metadata, \textit{direct} nodes, which point directly to data blocks, and \textit{indirect} nodes, which extend the addressability of large files beyond the capacity of direct nodes.
These blocks are grouped into \textit{segments}, which are further organized into \textit{sections}, the smallest unit that can be erased.
To support efficient block management and \textit{filesystem-level GC}, F2FS maintains several internal data structures:
(1) The \textit{Node Address Table (NAT)} maps node IDs to block addresses, (2) The \textit{Segment Information Table (SIT)} records metadata such as the number of valid blocks per segment and validity bitmaps, and (3) The \textit{Segment Summary Area (SSA)} logs summary information for each written block, including the owning inode number and logical offset, to aid in GC and recovery.
By adopting and customizing these structures, \name enables efficient metadata lookup, scalable file management, and flash-aware block allocation, while also paving the way for tight integration with the FTL.

\Cref{fig:D2} illustrates the overall architecture of \name. 
\name partitions the storage space into segments and employs a log‑structured layout, writing data sequentially across multiple segments.
For ease of management, \name sets the segment and section sizes to match the superblock size, which serves as the unit of erasure and GC in NAND flash.
\name maintains separate metadata and data sections similar to F2FS.
Accordingly, the metadata region and data region are logically allocated and separated at superblock granularity.

\name incorporates streamlined versions of key metadata structures from F2FS, adapted for its in-storage deployment.
First, it introduces a \textit{File Map}, which serves a role similar to F2FS's NAT but maps each node ID to its corresponding PPA  in flash. 
It also retains the \textit{SIT} to track valid \textit{flash} pages within each segment.
Both File Map and SIT reside in the reserved NAND region conventionally used for L2P mappings and SSD system metadata and are loaded into the device's onboard DRAM at boot time or during runtime.
To support GC and recovery, \name embeds lightweight summary information, such as the parent node ID and logical offset, in the OOB area of each flash page, eliminating the need for a dedicated SSA structure in DRAM.
\subsection{Co-design of Filesystem and FTL}\label{codesign}

To achieve high resource efficiency, \name is co-designed with the FTL.
Rather than treating the filesystem and FTL as independent layers, \name integrates their responsibilities by aligning metadata structures and eliminating redundant functionality.
This integration begins with the observation that F2FS's layout naturally reflects the characteristics of NAND flash: node and data blocks align with flash pages, and sections correspond to \textit{superblocks}.
\name structures its metadata to reflect these physical properties, streamlining coordination with the underlying FTL.
\begin{figure}[t]
    \centering
    \begin{subfigure}[t]{0.39\columnwidth} 
        \centering
        \includegraphics[width=\textwidth]{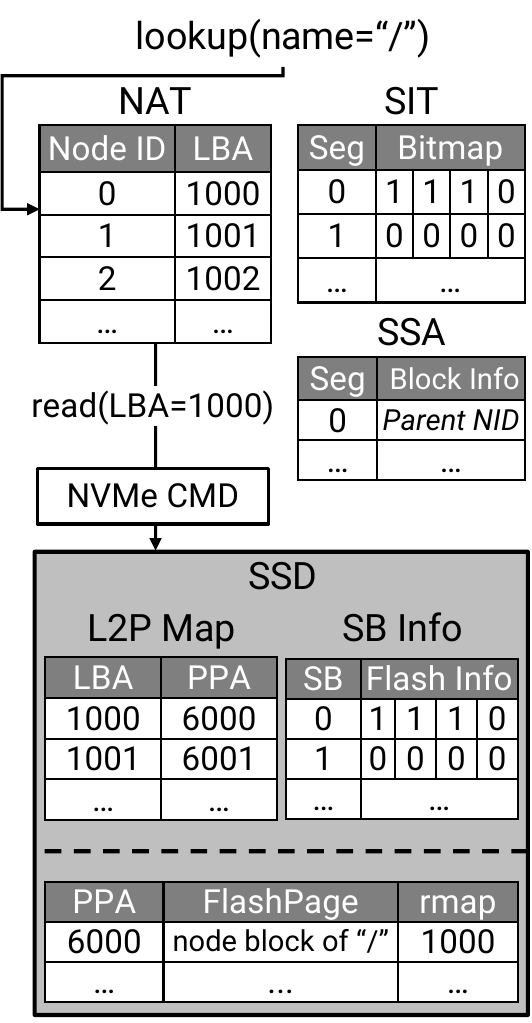} 
        \caption{F2FS} 
        \label{rfig:R1-1}
    \end{subfigure}
    \hspace{0.5cm}
    \begin{subfigure}[t]{0.378\columnwidth} 
        \centering
        \includegraphics[width=\textwidth]{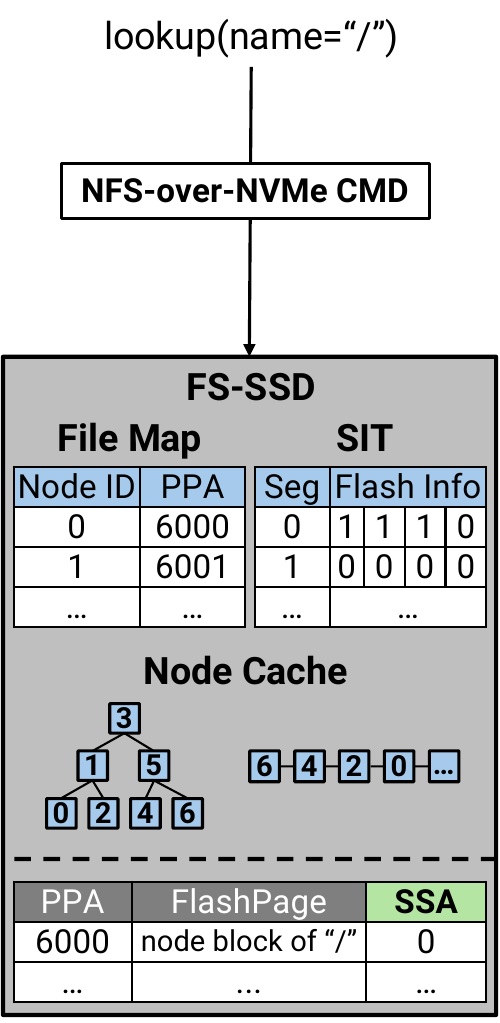} 
        \caption{\name} 
        \label{rfig:R1-2}
    \end{subfigure}
    \vspace{-0.3cm}
    \caption{\textbf{Comparison of metadata structures between F2FS and \name.}} 
    \vspace{-0.3cm}
    \label{rfig:R1}
\end{figure}
\Cref{rfig:R1} illustrates a high-level comparison of the metadata structures between F2FS and \name.
In F2FS, the NAT maps node IDs (NIDs) to LBAs, which are subsequently translated into PPAs by the FTL.
For example, when accessing the root directory (NID = 0), the NAT maps it to LBA 1000, and the host issues an NVMe \textsc{read} for that LBA.
Upon receiving the command, the SSD consults the L2P mapping table to locate the corresponding physical page at PPA 6000 and retrieves the node.

In contrast, \name integrates these two layers into a single File Map. 
When an application looks up the root directory, the node block can be retrieved directly using the \proto \textsc{lookup} command, allowing \name to translate the NID to its corresponding PPA through the File Map.
This mapping integration eliminates the need for separate L2P mappings, significantly reducing the DRAM footprint within the device.
Similarly, traditional FTLs maintain system metadata in the device DRAM for each superblock, including a validity bitmap to identify valid pages that must be relocated during GC (\textit{SB Info} in~\Cref{rfig:R1-1}).
In \name, this functionality is effectively integrated with the SIT by aligning the segment size with the superblock, allowing the status of node and data blocks to directly correspond to that of physical pages.

In conventional SSDs, the FTL requires \textit{reverse mapping (rmap)} information for each physical page to update the corresponding L2P mapping entry when a block is relocated during GC.
Typically, this \textit{Logical Page Number (LPN)} information is recorded in the OOB area of the flash page.
F2FS requires similar metadata to support its filesystem-level GC; it maintains the SSA that records the parent NID and offset in the parent node, which are used to update references in the node blocks in case the data block is moved during GC.
Since \name has no separate L2P mapping table, these metadata can be combined by writing the SSA entry directly into the OOB region during NAND programming.
As a result, \name avoids the need for a dedicated SSA and incurs no additional DRAM space to maintain this information.
Through these cross-layer metadata integrations between the filesystem and FTL, \name can natively support file abstractions within the same DRAM budget as conventional SSDs.

\subsection{Unified Garbage Collection}\label{unified_gc}
As data overwrites invalidate previously written blocks, F2FS performs GC at the filesystem level, similar to the GC mechanism employed by the FTL.
However, when F2FS runs atop a conventional SSD, the device has no visibility into these invalidations.
Flash pages corresponding to the invalidated LBAs are not explicitly marked as unused on the device. 
Consequently, filesystem- and FTL-level GC operate independently without coordination, leading to redundant data migrations and an increased Write Amplification Factor (WAF).
In contrast, \name directly manages the physical page states through its File Map and node structures. 
Data invalidation in \name immediately marks the corresponding flash pages as obsolete, effectively acting as a built-in \textsc{TRIM} command.
Thus, filesystem-level GC in \name inherently performs FTL-level GC, unifying what are traditionally separate processes. 
This unified GC eliminates redundant copying and enables \name to achieve a lower WAF than conventional host-managed filesystems.

To further enhance GC efficiency, \name supports file-based stream isolation, enabled by its ability to directly manage file-level metadata within the storage device. 
Users can define their own hot and cold data streams by specifying a special attribute in the inode. 
Since each stream is allocated to separate superblocks, \name maintains isolation between streams during GC, thereby preventing data intermixing.
This mechanism allows users to segregate specific files from others similar to the Flexible Data Placement (FDP) SSDs~\cite{fdp}.


\subsection{Crash Consistency}\label{crash_consistency}

F2FS ensures crash consistency through a checkpointing mechanism that periodically records a consistent snapshot of metadata.
This involves constructing a checkpoint pack that includes modified metadata.
During checkpointing, metadata updates are paused while the pack is flushed to flash. 
Upon recovery, F2FS loads the latest checkpoint pack and replays subsequent updates in a roll-forward manner~\cite{f2fs}.

Compared to F2FS, \name{} adopts simplified crash consistency inspired by DevFS~\cite{devfs}.
It uses power loss protection (PLP) capacitors commonly available in modern enterprise SSDs~\cite{PM893, 7450pro}.
These capacitors provide sufficient backup power to flush volatile metadata from DRAM to NAND flash during unexpected power failures. 
By using these capacitors, \name eliminates the need for checkpoint packs and avoids interrupting metadata updates.
Instead, \name maintains a log of in-flight requests in DRAM to safeguard against power failures during filesystem operations that modify internal states. 
Each log entry contains the opcode of an ongoing file operation in \name along with parameters, such as the file name and file handle of the parent directory. 
It also records phase information indicating how far the operation has progressed. 
For example, the \textsc{create} operation consists of multiple phases, depending on whether metadata access is required: (1) validating the parent directory’s dentry, (2) allocating a new NID and inode, and (3) inserting a new dentry into the parent directory.
During execution, \name records the completion of each phase along with the corresponding modification details in the log entry.

If a sudden power loss occurs, \name ensures durability by flushing the operation log, write buffer, and dirty metadata.
Upon system reboot, \name examines the log to detect any unfinished requests and replays the remaining phases of each interrupted operation to restore a consistent state, following an approach similar to F2FS's roll-forward mechanism.
To verify the correctness of this scheme, we conduct fault-injection tests 100 times while running synthetic workloads with eight threads on \name that update the filesystem state such as repeated file creation and deletion. 
We confirm that \name correctly identifies the phase at which the fault occurred and successfully recovers to a consistent state.

Our current design assumes that the aggregate size of this state remains within the device's PLP flush budget.
To reduce the amount of metadata that must be persisted upon power loss, \name could be extended to flush dirty metadata to NAND when its size reaches a predefined threshold, following the same approach used by conventional SSDs to bound dirty L2P updates in DRAM~\cite{spartanssd, lexar}.



\subsection{Concurrency Control}
\name basically uses node blocks as the granularity of locking to coordinate concurrent filesystem operations. 
Each metadata operation acquires locks on all node blocks it modifies. 
For example, \proto operations that modify the same directory such as \textsc{create} or \textsc{remove} are serialized through the lock on its parent-directory inode node, protecting shared directory state across requests from different clients and submission queues.
Operations spanning multiple directories require coordinated locking. 
For a cross-directory rename, \name acquires the locks on both parent-directory nodes in increasing NID order to prevent deadlock. 
It holds both locks until the source and destination namespace updates are complete, preventing conflicting metadata operations from interleaving with these updates. 

GC follows the same synchronization scheme since its unified mechanism operates with File Map and node blocks. 
Before relocating data pages, GC acquires the lock on their owner node, coordinating relocation with foreground operations that modify the same node. 
This synchronizes background reclamation with foreground metadata updates to shared file mappings.

\subsection{DRAM Resource Management}\label{device_resource}

Typically, the onboard DRAM in commodity SSDs accounts for only 0.1\% of the total storage capacity~\cite{970pro,990pro}. 
For example, 1\,TB SSDs are equipped with 1\,GB device DRAM, with most of it consumed by the FTL for maintaining the L2P mapping table.
We assume that \name is equipped with the same ratio of device DRAM for filesystem metadata. 

Although \name integrates the filesystem layer into the traditional FTL functionalities, it is carefully engineered to operate within these DRAM constraints.
\name maintains the File Map and SIT in DRAM, while allocating the remaining space to a \textit{node cache} that accelerates path traversal and metadata lookups as shown in~\Cref{rfig:R1-2}.
For example, with a 1\,TB \name and its 1\,GB DRAM, it reserves 2\% of the DRAM (i.e., 20\,MB) for File Map entries at startup, which is sufficient to index approximately five million node blocks. 
As the number of files grows, the File Map expands in 4~KB increments by reclaiming memory from the node cache.
In the case of SIT, it requires one bit per 4~KB block for its validity bitmap, which is 32~MB on a 1\,TB \name, so it occupies 3\% of the DRAM.
While the size of SIT increases linearly with the SSD capacity, this ratio remains consistent because the size of \name's DRAM also scales proportionally with its capacity.  
Thus, on average, around 95\% of the DRAM remains available for the cache.
The node cache stores 4\,KB node blocks organized using a red-black tree with an LRU eviction policy.
Since direct and indirect nodes--functionally equivalent to L2P mappings--require DRAM proportional to 0.1\% of the occupied NAND space, the node cache can, in the best case, cover up to 95\% of the addressable disk space while still meeting the tight memory budgets of modern SSDs.

\subsection{Implementation}
\noindent \textbf{Device Emulation.} 
We implement \name by extending NVMeVirt~\cite{nvmevirt, empowering}, a state-of-the-art software-defined NVMe device. 
NVMeVirt emulates the behavior of an NVMe device at the PCIe layer and faithfully models the performance characteristics of modern SSDs. 
Prior works have demonstrated that NVMeVirt enables detailed performance analysis based on actual I/O operations~\cite{aint, d2fs, order, sode}. 
Among the available performance models, we use the Samsung 970 Pro configuration~\cite{970pro}, which accurately replicates complex FTL behavior including contention across PCIe and NAND channels.

Upon initialization, NVMeVirt spawns two types of threads: a dispatcher and an I/O worker. 
The dispatcher continuously monitors the SQ for incoming NVMe commands.
When a command is issued, the dispatcher dequeues it and estimates its completion time based on the current state of in-flight operations and channel contention before assigning it to an I/O worker.
The I/O worker processes the request using reserved backend memory and posts a completion entry to the CQ once the predetermined completion time is reached.

\name is implemented within the I/O worker logic of NVMeVirt.
When an \proto command is received, the dispatcher immediately forwards the request to an I/O worker without precomputing the completion time.
This is because \name requires internal operations, such as file map and metadata lookup, prior to accessing flash pages.
The I/O worker measures the time spent executing these operations at runtime and includes it in the overall completion latency.
It allows \name to accurately reflect the in-storage processing cost and enables realistic end-to-end performance evaluation.
We deploy eight I/O workers on CPU cores configured to run at 1.4\,GHz, thereby emulating eight embedded cores.
%

\begin{figure}[t]
    \centering
    \begin{subfigure}[ht]{0.32\columnwidth} 
        \centering
        \includegraphics[width=\textwidth]{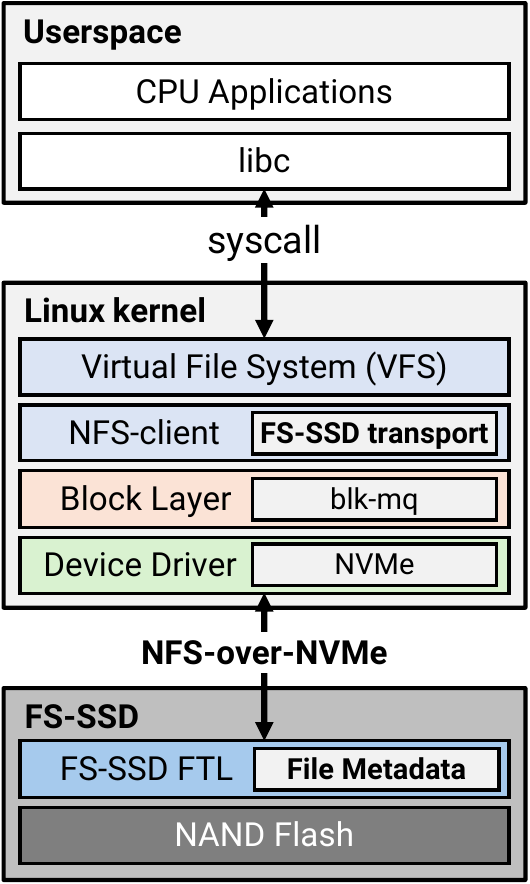} 
        \caption{\name-K} 
        \label{fig:R2-1}
    \end{subfigure}
    \hfill
    \begin{subfigure}[ht]{0.32\columnwidth}
        \centering
        \includegraphics[width=\textwidth]{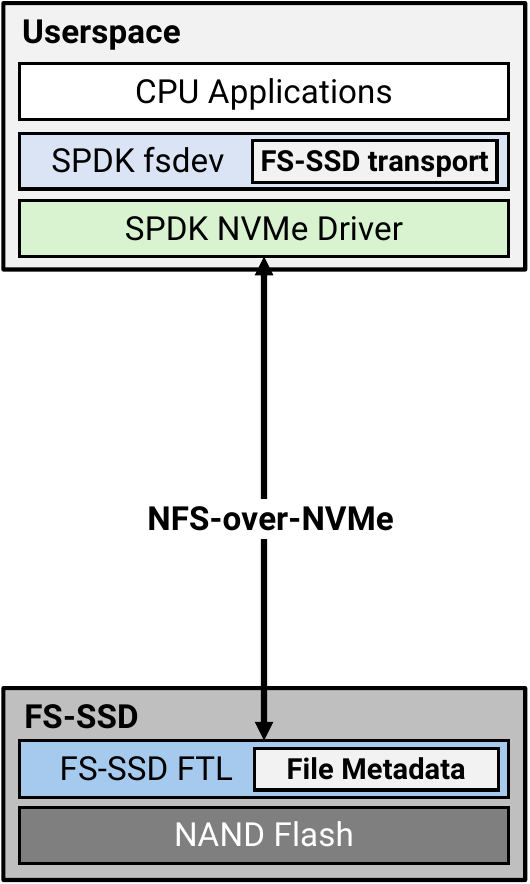} 
        \caption{\name-U} 
        \label{fig:R2-2}
    \end{subfigure}
    \hfill
    \begin{subfigure}[ht]{0.32\columnwidth} 
        \centering
        \includegraphics[width=\textwidth]{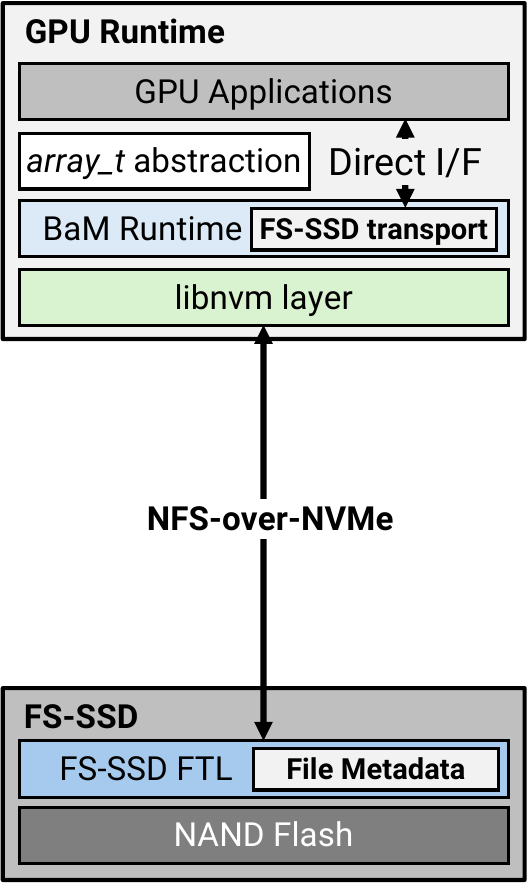} 
        \caption{\name-G} 
        \label{fig:R2-3}
    \end{subfigure}
    \vspace{-0.3cm}
    \caption{\textbf{\name client stacks.}}
    \vspace{-0.5cm}
    \label{fig:R2}
\end{figure}

\noindent \textbf{Client Implementations.} 
Since \name natively provides file abstraction within the device, it can support various heterogeneous client implementations.
As shown in~\Cref{fig:R2}, we implement three types of \name clients. 
We include further details of the \name clients in the Appendix \textcolor{ACMPurple}{A}.

(1) In-kernel client (\name-K): 
It is integrated into the in-kernel NFS client implementation by replacing its RPC transport layer to directly handle \proto commands. 
Because \name-K is managed by the VFS layer, it is exposed as a standard filesystem mount point to applications.

(2) Userspace client (\name-U): 
It is implemented as a usersp-ace direct client based on SPDK fsdev~\cite{spdk,fsdev}, thereby bypassing the kernel I/O stack for filesystem operations. 
Specifically, \name-U provides an fsdev backend that translates filesystem operations issued by SPDK applications into \proto commands.

(3) GPU-side client (\name-G): 
It is implemented by extending the BaM framework's I/O stack, enabling GPU threads to directly initiate file I/O to a \name device.
\name-G exposes two interfaces for accessing file data stored in the \name device: \textit{Direct File Interface (Direct I/F)} and extended \texttt{array\_t} \textit{abstraction}.
With the Direct I/F, a GPU kernel directly interacts with the \name transport layer to construct \proto commands and submit them through GPU-side NVMe queue pairs managed by libnvm~\cite{smartio}.
With extended \texttt{array\_t}, it is initialized with a filename and can map array offsets to file offsets, providing an abstraction analogous to \texttt{mmap()} for GPU-side file access. 

\noindent\textbf{Consistency Model}
By default, \name client implementations use direct I/O semantics for coherent file sharing across heterogeneous clients.
This avoids stale data caused by client-side caches by ensuring that reads and writes are served by \name rather than by cached host pages.
Thus, \name centralizes file metadata, access checks, and data placement inside the SSD while allowing each client to use an I/O path tailored to its execution environment.
When client-side caching is enabled, such as \name-K with the page cache, \name follows the \textit{close-to-open consistency}~\cite{linuxnfs} commonly used by NFS clients.
Under this model, updates made by one client become visible to another client after the writer closes the file and the reader subsequently opens it.
\section{Evaluation}\label{evaluation}

\subsection{Experimental Setup}\label{expriment_setup}
\noindent \textbf{Testbed.} 
We use two types of testbeds for our experiments. 
The primary testbed is equipped with two Intel\textregistered{} Xeon\textregistered{} Gold 5218R CPUs, 256~GB of DDR4 memory, a ConnectX-5 NIC, and an NVIDIA A100 PCIe (40~GB). 
Unless otherwise specified, all experiments are conducted on this testbed.
The second testbed is equipped with two Intel\textregistered{} Xeon\textregistered{} Silver 4316 CPUs and 512~GB of DDR4 memory, along with a ConnectX-6 NIC.
For experiments in remote storage environments, the first server is used as the client, and the second server as the storage server.

\noindent \textbf{Software Setup.} 
Both testbeds run Ubuntu 20.04 LTS with Linux kernel version 6.2.16.
We use NVMeVirt~\cite{nvmevirt} configured with the Samsung 970 Pro model~\cite{970pro} as the backend storage device for all filesystems.
Half of each testbed's DRAM is allocated to NVMeVirt, while the remaining memory is reserved for the host system.
We provision on-device DRAM as 0.1\% of the total storage capacity and constrain the memory footprint of both the baselines and \name to this budget.
We use SPDK version 26.01 and its fsdev framework for \name-U.
For NFS and NFS-over-RDMA (NFS-oR), we use VAST NFS~\cite{vastnfs} version 4.0.32, which includes GDS support on the NFS client side.
For GPU-based experiments, we use GDS with nvidia-fs~\cite{nvidia-fs} version 2.17.4 and the latest BaM available on GitHub.
%

%
\subsection{Local Environment}\label{micro}

\begin{figure}[t]
    \centering
    \begin{subfigure}[t]{0.49\columnwidth}
            \includegraphics[width=\textwidth]{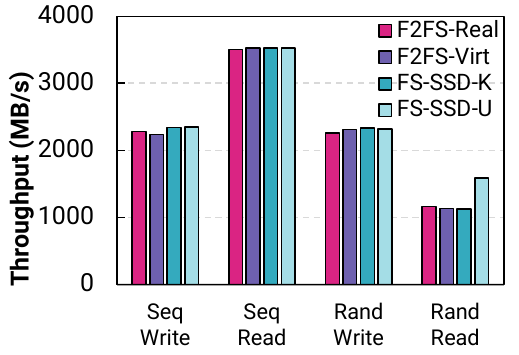} 
        \caption{\texttt{fio} throughput}
        \label{fig:E1} 
    \end{subfigure}
    \hfill
    \begin{subfigure}[t]{0.49\columnwidth}
            \includegraphics[width=\textwidth]{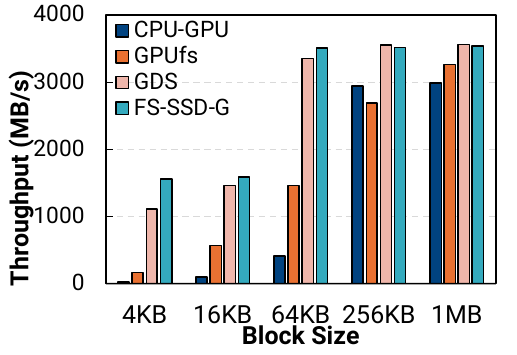} 
        \caption{I/O-centric throughput}
        \label{fig:E5}
    \end{subfigure}
    \vspace{-0.2cm}
    \caption{\textbf{I/O Performance in a local environment.}}
    \vspace{-0.4cm}
\end{figure}
%
\noindent \textbf{Baseline Throughput.} 
We evaluate the impact of \name's design and NFSv3 semantics on I/O performance by running \texttt{fio} on F2FS, \name-K and \name-U in a local deployment.
We also report F2FS performance on a real 970 Pro SSD to provide a reference for our NVMeVirt-based results.
Filesystems labeled with \textit{Real} denote results on the physical device, while \textit{Virt} indicates results from NVMeVirt.
Sequential I/O tests use a single thread with 128\,KB requests, whereas random I/O tests use 32 threads with 4\,KB requests.
All benchmarks are executed using the \textit{libaio} ioengine with the \textit{direct} option enabled and a queue depth of 128.
For \name-U, we use a \texttt{fio} fsdev ioengine implemented by extending the existing \texttt{fio} bdev ioengine~\cite{bdev}.

\Cref{fig:E1} presents the throughput results for each configuration.
We can see that all configurations saturate the underlying device for both sequential reads and writes, closely matching the performance of the real device across different storage access patterns.
Random write throughput is comparable to sequential write, primarily due to the log-structured designs of F2FS and \name. 
In contrast, for random reads, \name-U achieves 41\% higher throughput than the others owing to its kernel-bypass file I/O path based on SPDK fsdev.
In summary, we observe that \name's F2FS-based design and NFSv3 semantics incur no significant performance overhead compared to in-kernel F2FS and fully utilize the bandwidth of a raw device while offering on-device file abstraction.


\noindent \textbf{Throughput in I/O-centric Architectures.} 
We evaluate the effectiveness of \name in I/O-centric approaches by measuring the throughput of data-transfer scenarios\footnote{We omit state-of-the-art coordinated filesystems such as GeminiFS and GoFS from the comparison because their codes are not publicly available.} between a GPU and an SSD in local environments. 
Specifically, we evaluate: (1) the CPU-GPU approach using \texttt{cudaMemcpy()}, (2) GPUfs~\cite{gpufs}, (3) GDS, and (4) \name-G.
GPUfs is a GPU-side library that enables GPU kernels to directly invoke filesystem APIs.
When kernels issue API calls, GPUfs delegates the corresponding operations to the host through RPCs and the host CPU executes the requested filesystem operations.
Subsequently, data is transferred between the CPU and GPU using \texttt{cudaMemcpy()}.
The CPU-GPU and GPUfs use F2FS as the underlying host filesystem, whereas GDS uses EXT4 because it does not support F2FS as a backend filesystem.

We run random read workloads across these scenarios with block sizes ranging from 4\,KB to 1\,MB, and the results are shown in~\Cref{fig:E5}.
We observe that \name-G outperforms the baselines across all block sizes while leveraging \name's on-device file abstraction without involving the host CPU in the file I/O path.
Although GPUfs achieves higher throughput than CPU-GPU for small block sizes by allowing GPU kernels to issue file I/O directly, data transfers through \texttt{cudaMemcpy()} still become a bottleneck.
GDS alleviates this overhead by enabling direct data transfers between storage and GPU memory, but it still exhibits up to a 40\% throughput gap for block sizes smaller than 64\,KB because it relies on the host filesystem to locate data.

\begin{figure}[t]
    \centering
    \begin{subfigure}[t]{0.49\columnwidth}
            \includegraphics[width=\textwidth]{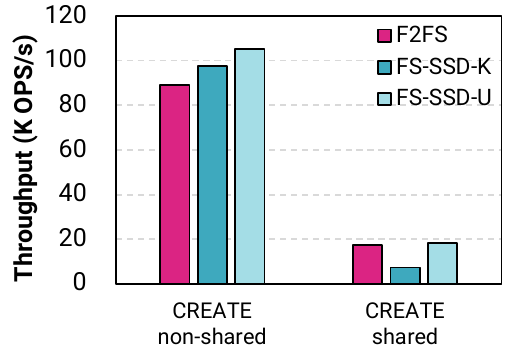} 
        \caption{File creation}
        \label{fig:E9-1} 
    \end{subfigure}
    \hfill
    \begin{subfigure}[t]{0.49\columnwidth}
            \includegraphics[width=\textwidth]{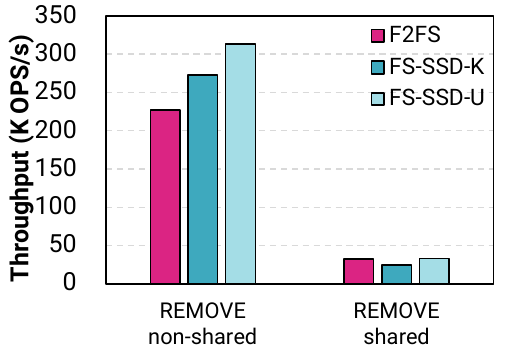} 
        \caption{File deletion}
        \label{fig:E9-2}
    \end{subfigure}
    \vspace{-0.25cm}
    \caption{\textbf{Metadata Performance.}}
    \Description{.}
    \vspace{-0.45cm}
    \label{fig:E9}
\end{figure}

\noindent \textbf{Garbage Collection with Stream Isolation.} 
To evaluate the impact of unified GC in \name, we generate a synthetic workload using the \texttt{fio}.
This workload consists of eight threads: four threads performing sequential writes and the other four threads performing random writes.
We run this workload on both F2FS and \name, and measure the Write Amplification Factor (WAF), defined as the ratio of the total amount of data written to flash pages over the total amount of data written by the application.
We also evaluate \name with the stream isolation mechanism (\name-stream), where the sequentially written file is assigned to a cold stream. 

We observe WAF of 4.22 for F2FS, whereas \name achieves a 63\% lower WAF of 1.53 by unifying filesystem- and FTL-level GC, eliminating redundancy and reducing unnecessary data movement.
\name-stream further reduces WAF to 1.30 by physically isolating flash pages belonging to sequentially written files from other data.
This allows GC to be confined to flash blocks containing randomly written data, thereby reducing the volume of valid pages that must be migrated. 

\noindent \textbf{Metadata Operation Throughput.}
We measure the metadata operation performance of \name using file creation and deletion workloads and compare its throughput with that of F2FS.
For the deletion workload, we first create files until the filesystem returns \texttt{ENOSPC}.
We run each workload for 30 seconds with 16 threads under two conditions: \textit{non-shared}, where each thread operates in a separate directory, and \textit{shared}, where all threads operate in a shared directory.

As shown in~\Cref{fig:E9}, both \name-K and \name-U achieve higher throughput than F2FS under the non-shared condition.
In F2FS, metadata-intensive workloads cause dirty node pages and NAT entries to accumulate beyond their thresholds, triggering checkpointing that flushes dirty metadata through block-level I/O and temporarily stalls foreground metadata operations~\cite{f2fs, f2fsj}.
In contrast, \name performs the operations within the device by sending \proto commands and protects them using its lightweight operation log, reducing host-side metadata management overhead.

Under the shared condition, however, \name-K achieves lower throughput than F2FS.
This result is imposed by the inode semaphore (\texttt{i\_rwsem}) in the VFS layer.
Although \name can coordinate concurrent directory operations within the device, the VFS layer acquires this semaphore exclusively for file creation and deletion.
In \name-K, the semaphore remains held until the device request completes.
Thus, the critical section additionally includes PCIe communication latency in addition to metadata processing time, increasing lock contention and limiting throughput.
By bypassing the semaphore, \name-U avoids this additional serialization and achieves throughput comparable to that of F2FS, suggesting that the throughput is ultimately limited by contention on shared directory's metadata in the device.

\subsection{Remote Environment}\label{remote}

\noindent \textbf{Baseline Throughput.} 
\name can support direct access from remote clients without kernel intervention through the \proto protocol.
The commands are transmitted to the remote server via RDMA and the remote server's NIC forwards them directly to the \name, bypassing the remote CPU and kernel.
To evaluate its effectiveness, we compare \name against NFS and NFS-oR, both using F2FS as their backend filesystem.
In contrast to \name, NFS relies on RPC-based communication, where each request is handled by the NFS server, passed through the VFS layer, and then issued to the storage stack, incurring CPU and kernel involvement.
NFS-oR follows the same execution path as NFS, but uses RDMA as the RPC transport to reduce network overhead.

\begin{figure}[t]
    \begin{subfigure}[t]{0.49\columnwidth}
        \centering
            \includegraphics[width=\textwidth]{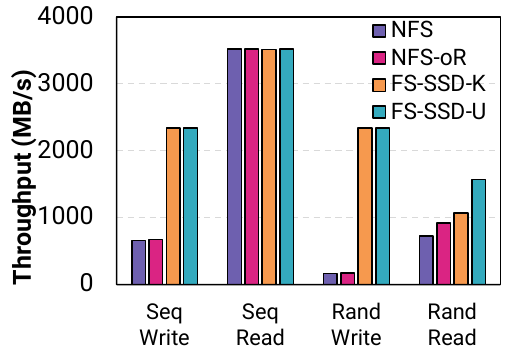} 
        \caption{\texttt{fio} throughput}
        \label{fig:E4-1} 
    \end{subfigure}
    \hfill
    \begin{subfigure}[t]{0.49\columnwidth}
        \centering
            \includegraphics[width=\textwidth]{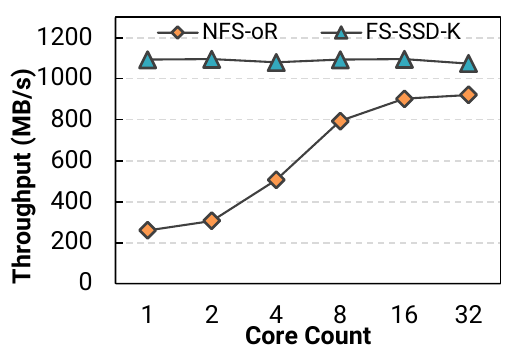} 
        \caption{Impact of target core count} 
        \label{fig:E4-3}
    \end{subfigure}
    \vspace{-0.3cm}
    \caption{\textbf{I/O Performance in a remote environment.}}
    \vspace{-0.3cm}
\end{figure}

We run the \texttt{fio} using the same workloads described in \Cref{micro} and present the results in \Cref{fig:E4-1}. 
Overall, \name-K and \name-U consistently outperform both NFS and NFS-oR across all workloads. 
Notably, the write throughput of NFS and NFS-oR is significantly lower than other configurations, primarily due to their deep software stack, which incurs multiple memory copies and context switches.
In contrast, \name sustains throughput comparable to local environments, demonstrating its ability to maintain high efficiency even in remote environments.

\noindent \textbf{Target CPU Usage.} 
To further examine the impact of host CPU usage on the performance of \name and NFS-oR, we run a 4~KB random read workload while varying the number of CPU cores on the target server from 1 to 32. 
As shown in \Cref{fig:E4-3}, NFS-oR throughput is severely constrained by the number of available target cores, dropping by 76\% compared to \name when only a single core is available.
On the other hand, \name maintains consistent throughput regardless of the number of host CPU cores. 
This is because \name fully bypasses the host CPU of target server via Target Offload, which allows the NIC to forward \proto commands directly to the storage device. 
In contrast, NFS-oR incurs high CPU utilization due to its reliance on the host kernel and VFS layer.
NFS-oR reduces CPU usage by avoiding redundant memory copies via RDMA, but it still consumes considerable CPU resources since filesystem operations are still handled in software on the host.
We argue that \name's minimal reliance on CPU resources not only improves scalability but also has the potential to reduce the total cost of ownership (TCO) for deploying high-performance remote file services.

\subsection{Real-world I/O-centric Workloads}\label{real-app}
We assess \name's capability to provide a scalable and efficient file abstraction for GPU-centric, I/O-intensive architectures by running two representative GPU workloads.

\noindent \textbf{Graph Analytics.} 
We evaluate the benefits of \name-G using two graph analytics workloads: Breadth-First Search (BFS) and Connected Components (CC). 
The experiments are conducted in a local environment using the three graph datasets listed in~\Cref{tab:graphs}. 
We exclude \texttt{uk} from the CC experiments because CC assumes an undirected graph. 
Across the datasets, we measure the total execution time of each workload on BaM (labeled as \texttt{B}) and \name-G (labeled as \texttt{FG}). 
BaM runs on NVMeVirt configured with the Samsung 970 Pro performance model, which exposes only a raw block interface to BaM.
Thus, BaM requires a \textit{Staging} phase that preloads graph datasets into fixed SSD locations before execution.
To reduce variation from reading the source datasets during the staging phase, we place the source dataset files in DRAM before preloading them onto the SSD.

\begin{figure}[t]
    \centering
    \begin{subfigure}[t]{0.485\columnwidth}
        \centering
        \includegraphics[width=\textwidth]{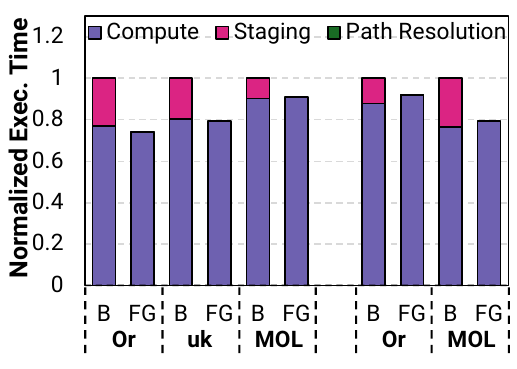} 
        \caption{Graph analytics} 
        \label{fig:E7-3}
    \end{subfigure}
    \hfill
    \begin{subfigure}[t]{0.49\columnwidth}
        \centering
        \includegraphics[width=\textwidth]{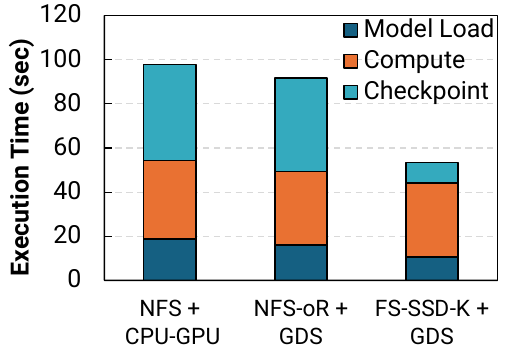} 
        \caption{GPT3-1.3B training} 
        \label{fig:E7-2}
    \end{subfigure}
    \vspace{-0.3cm}
    \caption{\textbf{Performance of GPU workloads.}}
    \vspace{-0.6cm}
    \label{fig:E7}
\end{figure}

As shown in \Cref{fig:E7-3}, BaM incurs substantially longer execution time than \name-G on average. 
Specifically, BaM spends up to 25\% of its total execution time in the staging phase.
In contrast, with \name-G, datasets need not be staged at fixed SSD offsets, nor does the runtime need to know their block-level locations.
Instead, \name-G replaces BaM's staging phase with lightweight \textit{Path Resolution} that resolves a target filename to a file handle, which accounts for only a negligible fraction of the total execution time.
Subsequent I/Os are issued using the file handle and file offsets, while \name locates the data internally.
The results show that \name-G improves not only the usability of I/O-centric architectures, but also their performance.


\begin{table}[t]
    \centering
    \caption{\textbf{Graph analytics datasets.}}
    \vspace{-0.3cm}
    \resizebox{0.9\columnwidth}{!}{%
    \begin{tabular}{cccc}
        \Xhline{3\arrayrulewidth}
        \textbf{Graph} & \textbf{\# Nodes} & \textbf{\# Edges} & \textbf{File Size} \\ \Xhline{3\arrayrulewidth}
        \texttt{Orkut (Or)}~\cite{orkut}           & 3.1\,M    & 234.3\,M  & 1.7\,GiB \\
        \texttt{uk-2007-05 (uk)}~\cite{uk}         & 105.9\,M  & 3.37\,B   & 66\,GiB \\
        \texttt{MOLIERE\_2016 (MOL)}~\cite{moliere} & 30.2\,M   & 6.67\,B   & 83\,GiB  \\
        \Xhline{3\arrayrulewidth}
    \end{tabular}%
    }
    \label{tab:graphs}
    \vspace{-0.4cm}
\end{table}

\noindent \textbf{LLM Training from a Checkpoint.} 
We perform LLM training using the GPT3-1.3B model, focusing on the scenario where training resumes from a checkpoint stored on remote storage.
After the \textit{Model Load} phase with the prepared model, each training step consists of two phases: \textit{Compute}, which updates model weights, and \textit{Checkpoint}, which writes updated weights and optimizer states to storage. 
The total checkpoint size for GPT3-1.3B comprises 2.5\,GB of model weights and 15~GB of AdamW optimizer state~\cite{adamw}.
Since each step consists of comparable compute and checkpoint phases, we report the execution time breakdown of the first step.
We evaluate remote storage under three setups: (1) NFS using the CPU-GPU, (2) NFS-oR with GDS, and (3) \name-K with GDS.
Since \name-G currently lacks remote support due to limitations of the underlying BaM runtime, we use GDS with \name-K to enable GPU access to a remote \name.

\Cref{fig:E7-2} demonstrates that \name achieves the lowest execution time across all configurations.
In particular, \name reduces checkpoint time by 78\% compared to NFS and 77\% compared to NFS-oR.
Notably, NFS-oR fails to fully leverage GDS's benefits due to the deep I/O stack in conventional remote NFS servers.
By contrast, as observed in \Cref{remote}, \name bypasses this remote I/O stack, enabling a more direct data path and achieving higher throughput, thereby reducing end-to-end execution time.
\section{Future Work}\label{discussion}

Although \name currently operates as a single-device storage solution, we regard this work as a foundational step toward building a distributed storage array with file semantics.
We are actively extending \name into a scalable, multi-device architecture based on the pNFS protocol introduced in NFSv4.1~\cite{NFSv4.1}, which inherently supports parallel data access and distributed metadata coordination with NFSv4-compliant file locking semantics. 
Our ongoing work envisions a distributed \name system that delivers high performance through a fully hostless file I/O path to the storage array, sustaining \name's resource efficiency and requiring only minimal host involvement for array management.
This design effectively reduces the TCO while preserving the simplicity and efficiency of \name's device-level file abstraction.
Ultimately, it aims to bridge the gap between \name and large-scale, software-defined storage infrastructures~\cite{3fs, lustre}.

\vspace{-0.1cm}
\section{Related Work}\label{related_work}

\noindent\textbf{Computational Storage Device.} 
Computational Storage Device (CSD) embeds compute engines (e.g., ARM cores or FPGAs), enabling computations to execute near the data.
For example, Samsung SmartSSD Gen2~\cite{samsung_smartssd2.0} and Newport CSD~\cite{newportcsd} incorporate dedicated ARM cores to offload functions from the host CPU.
By reducing host CPU utilization and data movement, CSDs have been shown to benefit various data‑intensive workloads~\cite{lambda-io, presto, insider, smartanns, neodbms, summarizer}.
However, as noted in~\Cref{background}, CSDs lack an internal file abstraction and therefore rely on the host filesystem.
Unlike CSDs, \name supports on-device file access without redundant host-device round trips and additional computing resources thanks to a native file layer co-designed with the FTL.
With a modest increase in cores and DRAM, we believe \name can also support function offloading like CSDs while eliminating communication overhead to locate the file data.

\noindent\textbf{Object Storage Device.}
Object Storage Device (OSD) exposes data as variable-sized objects identified by unique object IDs rather than fixed-sized blocks, delegating storage allocation and placement to the device. 
The T10 OSD standard~\cite{t10} defines a SCSI command set for this abstraction, including object creation and removal. 
These operations allow clients to manipulate objects and their attributes without relying on the OS kernel to manage block-level placement.
Although an OSD may be simpler to implement than \name, it still requires an external layer to map application data to objects and coordinate concurrent updates of related objects.
In practice, applications must manage data-to-object ID mappings and coordinate related updates themselves, or delegate these responsibilities to an OSD-aware host filesystem such as exofs~\cite{osdfs, exofs}.
In contrast, \name moves responsibilities into the device, reducing the external coordination required for shared file access.
Furthermore, we argue that \name can be readily extended to serve as an OSD, if needed, by adding an object interface and using filenames as object IDs.

\noindent\textbf{Key-Value SSD.} 
Key-Value SSD (KVSSD) implements a key-value engine directly in the device firmware, allowing data to be managed via key-value pairs at the hardware level.
These devices support native KV operations such as \textsc{get}, \textsc{put}, and \textsc{delete}~\cite{nvme-kv}. 
By eliminating the traditional host KV software stack where key-based operations are translated into block-based operations, KVSSDs can significantly improve I/O performance for KV-based applications.
To manage key-indexed data, KVSSDs typically maintain an internal index implemented as a hash table~\cite{bluecache, nvmkv} or a log-structured merge (LSM) tree~\cite{kevin, ilsm-ssd}. 
For instance, Samsung KVSSD~\cite{kvssd} uses a multi-level hash table to enable fast point queries, whereas PinK~\cite{pink} adopts an LSM-tree to improve update costs and space efficiency. 
Although KVSSDs are viable options for I/O-centric architectures, their limited KV APIs hinder practical adoption.  
They lack built-in access control, complicating multi-tenant deployments. 
Moreover, writing large key-value data requires reserving substantial on-device buffers, as KV APIs lack fine-grained data flow control.

\vspace{-0.1cm}
\section{Conclusions}\label{conclusion}
In this paper, we propose \name, a new type of SSD that provides full-fledged file abstraction for I/O‑centric architectures while maximizing resource efficiency.
By adopting F2FS's flash-friendly layout and co-designing with the FTL, \name operates effectively within the limited DRAM and compute resources of modern SSDs.
We also introduce \proto, a novel protocol that enables direct file‑level access by peripheral devices without host CPU intervention. 
Combined with NVMe-oF Target Offload, NFS‑over‑NVMe allows \name to natively support remote storage access, completely bypassing the host I/O stack.





\bibliographystyle{ACM-Reference-Format}
\bibliography{references}

\clearpage
\newpage

\appendix
\section{\name Client Details}\label{clients}


\name supports three client implementations: an in-kernel client (\name-K), a userspace client (\name-U), and a GPU-side client (\name-G).
As shown in~\Cref{fig:R2}, these clients share the same device-side file interface, but differ in where \proto commands are generated and how they are submitted to \name.
This design allows \name to serve conventional CPU applications, kernel-bypass userspace applications, and GPU kernels through a native file abstraction.
The following subsections describe the design and implementation of each client I/O stack.

\subsection{\name-K: In-kernel Client}
\name-K provides a standard filesystem mount point for CPU applications.
It reuses the Linux in-kernel NFS client implementation, allowing \name-K to remain compatible with the Virtual File System (VFS) layer.
Specifically, we replace the NFS RPC transport layer, which sends NFS RPC messages through the Linux network stack, with the \textit{\name transport layer}.
This transport layer translates file operations into \proto commands and submits them to the \name device through the underlying block layer and NVMe device driver as illustrated in~\Cref{fig:R2-1}.

CPU applications can access \name through standard system calls issued via \textit{libc}.
Furthermore, this design enables \name to support GPUDirect Storage (GDS)~\cite{gds} as shown in~\Cref{real-app}.
When applications access files through the \textit{cuFile} interface of GDS, the nvidia-fs driver~\cite{nvidia-fs} orchestrates I/O directly between DMA-capable storage and user-allocated GPU memory.
Because the \name device supports DMA through its I/OAT DMA engine~\cite{ioat}, which was already integrated into the original NVMeVirt~\cite{nvmevirt}, \name-K can reuse the DMA callbacks already registered in nvidia-fs.
In particular, we register \name-K with \texttt{NVFS\_PROC\_MOD\_NFS\allowbreak\_KEY} while binding it to the DMA callbacks implemented for \texttt{NVFS\_PROC\_MOD\_NVME\_KEY}.
This allows file operations issued through cuFile to be translated into \proto commands, while the DMA target address in GPU memory is passed directly to the \name device.
As a result, \name can support GDS without significant modifications to the GDS stack.

\subsection{\name-U: Userspace Client}
\name-U implements a userspace file I/O stack for kernel-bypass file access to \name as demonstrated in~\Cref{fig:R2-2}.
It is implemented as an SPDK fsdev~\cite{fsdev} backend that exposes file operations to SPDK applications.
When an application invokes fsdev file APIs, \name-U generates the corresponding \proto commands through the userspace \name transport layer and submits them via userspace NVMe queue pairs managed by the SPDK NVMe driver.
Because both command submission and completion handling occur in userspace, \name-U avoids system calls on the file I/O path and uses SPDK's polling-based NVMe queue pairs to sustain high request rates.
This path is especially effective for small random I/O, as shown in Section~\textcolor{ACMPurple}{5.2}, because it bypasses the deep kernel software stack, including the VFS, block layer, and interrupt-driven completion path, while preserving the same file namespace as \name-K.

Consistent with SPDK's event-driven asynchronous design, SPDK fsdev exposes an asynchronous file I/O model that decouples request submission from completion through callback mechanisms. 
Therefore, if dependent file operations require the same synchronous ordering guarantees as \name-K, such ordering must be enforced at the SPDK application level.
This can be implemented with an explicit wait loop that waits for the previous completion callback before issuing the next operation, or with a callback chain that issues each subsequent operation from the completion callback of the preceding operation as recommended in the SPDK documentation~\cite{spdk-chain}.

\subsection{\name-G: GPU-side Client}
\name-G enables GPU kernels to initiate file I/O without involving the host CPU on the control or data path after initialization.
As illustrated in~\Cref{fig:R2-3}, it is implemented as a lightweight GPU-side runtime that extends the GPU-initiated I/O path of BaM~\cite{bam} with the \name transport layer.
\name-G exposes two interfaces for accessing file data stored in the \name device: 

(1) \textit{Direct File Interface (Direct I/F)}: With the Direct I/F, a GPU kernel directly interacts with the \name transport layer to construct \proto commands and submit them through GPU-side NVMe queue pairs managed by libnvm~\cite{smartio}.
The kernel then polls GPU-resident completion queues until the requested file operation completes.
We use the Direct I/F in the experiment shown in~Figure~\textcolor{ACMPurple}{8b} of the main manuscript.

(2) \texttt{array\_t} \textit{abstraction}: \texttt{array\_t} is BaM's high-level programming abstraction for creating an array backed by NVMe storage.
It allows GPU applications to access storage-resident data through an array-like interface.
Internally, BaM maintains a mapping from array offsets to blocks on NVMe SSDs and fetches data from the storage device on demand.
However, because each array entry is mapped directly to a fixed block location, the dataset must be preloaded onto the SSD before execution.
Otherwise, BaM cannot resolve array accesses to valid storage locations.

With \name, this redundant data staging can be eliminated because \name performs file-to-block translation inside the device.
To this end, we extend \texttt{array\_t} so that it can be initialized with a filename and can map array offsets to file offsets, providing an abstraction analogous to \texttt{mmap()} for GPU-side file access.

\begin{listing}[t]
\caption{Example using \texttt{array\_t} in \name-G.}
\label{lst:gpu-read}
\begin{minted}[
  linenos,
  breaklines,
  fontsize=\fontsize{7}{9}\selectfont,
  numbersep=-6pt
]{cuda}
   __global__
   void kernel(array_d_t<uint64_t> d_arr, size_t size) {
       uint64_t tid = blockDim.x * blockIdx.x + threadIdx.x;
       uint64_t stride = blockDim.x * gridDim.x;
       uint64_t i, data;
    
       for (i = tid; i < size; i += stride) {
           // Read file data on the GPU through the array abstraction
           data = d_arr[i];
           ...
       }
   }

   int main(void) { 
       ... 
       // Resolve the file name to a file handle in the constructor
       array_t<uint64_t> *arr = new array_t<uint64_t>(filename, ...);

       kernel<<<blockDim, numthreads>>>(arr->d_array, arr->size);
       ...
   }
\end{minted}
\end{listing}

Listing~\ref{lst:gpu-read} illustrates how users can leverage the extended \texttt{array\_t} abstraction in \name-G.
Before launching a GPU kernel, users create an \texttt{array\_t} instance by passing a filename to its constructor. 
The constructor internally resolves the filename to a file handle and prepares a GPU-side array instance (i.e., \texttt{array\_d\_t}).
Then, while the kernel is running, accesses to the array are translated into GPU-side file reads through the \name transport layer.
Thus, \texttt{array\_t} provides GPU applications with a higher-level file-backed array abstraction.
We use this \texttt{array\_t} abstraction in the experiment shown in~Figure~\textcolor{ACMPurple}{11a} of the main manuscript, where each graph dataset is represented as an \texttt{array\_t} instance.

\section{Impact of CPU core frequency}\label{cpu-freq}

Embedded CPUs in SSDs generally operate at lower frequencies than host CPUs, which is particularly important for \name since all filesystem operations are executed on the storage device. 
Since prior works~\cite{sode, kvftl} have shown that reducing the core frequency effectively emulates the characteristics of low-performance cores (e.g., ARM cores), we adopt the same approach to emulate wimpy device cores in our experiments.
This adjustment primarily affects internal operations such as file map lookup and node traversal during read requests.
To assess the impact of core speed on performance, we measure random read throughput while scaling the core frequencies of NVMeVirt's worker threads from 2.2\,GHz, which is near the maximum frequency of our host CPU, down to 1\,GHz, representing a low-performance device core.
All other cores remain at their default frequencies.
Notably, under this concurrent random-read workload, we observe that \name maintains stable throughput across the tested frequencies, while the average end-to-end request latency increases by only 2\% when the core frequency is reduced to 1\,GHz.

\begin{figure}[t]
    \centering
        \includegraphics[width=0.8\columnwidth]{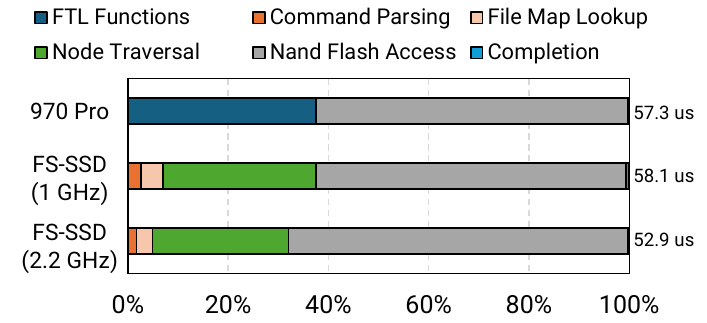} 
    \vspace{-0.2cm}
    \caption{\textbf{Normalized read latency breakdown.}}
    \label{fig:E8} 
    \vspace{-0.2cm}
\end{figure}

To further examine the impact of wimpy device cores on an individual request, we present the device-side read latency breakdown for two different frequency configurations on \name and compare them with NVMeVirt’s 970 Pro model, which accurately captures FTL firmware overheads on I/O operations.
In \Cref{fig:E8}, \name (1\,GHz) exhibits 9.8\% higher per-request latency than \name (2.2\,GHz) due to the reduced core frequency.
However, the latency of internal filesystem operations is relatively small and largely hidden by NAND flash access latency which accounts for approximately 70\% of the total. 
When I/Os are issued in parallel, these internal operations are effectively pipelined, resulting in only a negligible impact on the overall average latency.
Moreover, the latency of internal operations in \name (1\,GHz) is comparable to the FTL overheads in the 970 Pro, which indicates that the integrated filesystem operations are lightweight and efficient, making them well-suited for execution on resource-constrained storage devices.

\end{document}